\documentclass[prd,nofootinbib,showpacs,preprint]{revtex4-1}
\usepackage[T1]{fontenc}
\usepackage{amsmath,amssymb}
\usepackage{epsfig}
\usepackage{url}
\usepackage{graphicx}
\usepackage[usenames,dvipsnames]{color}
\usepackage{slashed}
\usepackage[colorlinks,citecolor=blue]{hyperref}
\usepackage{color}
\usepackage{cancel}

\usepackage{cancel}
 \usepackage[normalem]{ulem}
\usepackage{color}

\def\lsim{\;\raise0.3ex\hbox{$<$\kern-0.75em\raise-1.1ex\hbox{$\sim$}}\;}
\def\gsim{\;\raise0.3ex\hbox{$>$\kern-0.75em\raise-1.1ex\hbox{$\sim$}}\;}

\def\ben{\begin{enumerate}}  \def\een{\end{enumerate}}
\def\bit{\begin{itemize}}    \def\eit{\end{itemize}}
\def\beq{\begin{equation}}   \def\eeq{\end{equation}}
\def\ba{\begin{array}}       \def\ea{\end{array}}
\def\bea{\begin{eqnarray}}   \def\eea{\end{eqnarray}}

\newcommand{\cl}{{\cal{L}}}
\def \met  {\mbox{${E\!\!\!\!/_T}$}}

\newcommand{\comment}[1]{}

\begin{document}
\title{Probing sterile neutrinos in the framework of inverse seesaw mechanism through leptoquark productions}
\preprint{HRI-RECAPP-2017-011} 
\preprint{IP/BBSR/2017-10}
\author{Debottam Das}
\email{debottam@iopb.res.in}
\author{Kirtiman Ghosh}
\email{kirti.gh@gmail.com}
\affiliation{Institute of Physics, Bhubaneswar 751005, India \& Homi Bhabha National Institute, Training School Complex, Anushakti Nagar,
Mumbai 400085, India}
\author{Manimala Mitra}
\email{ manimala@iopb.res.in}
\affiliation{Institute of Physics, Bhubaneswar 751005, India \& Homi Bhabha National Institute, Training School Complex, Anushakti Nagar,
Mumbai 400085, India}
\author{Subhadeep Mondal}
\email{subhadeepmondal@hri.res.in}
\affiliation{Regional Centre for Accelerator-based Particle Physics, Harish-Chandra Research Institute, 
HBNI, Chhatnag Road, Jhunsi, Allahabad 211019, India \& Department of Physics, and Helsinki Institute of Physics, P. O. Box 64, FI-00014 University of Helsinki, Finland}
\begin{abstract}
We consider an extension of the Standard Model (SM) augmented by two neutral singlet fermions per generation  
and a leptoquark. In order to generate the light neutrino masses and mixing, we incorporate inverse seesaw 
mechanism. The right handed neutrino production in this model is significantly larger than the conventional 
inverse seesaw scenario. We analyze the different collider signatures of this model and find that the final 
states associated with three or more leptons, multi jet and at least one b-tagged and (or) $\tau$-tagged 
jet can probe larger RH neutrino mass scale. We have also proposed a same-sign dilepton signal region 
associated with multiple jets and missing energy that can be used to distinguish the the present scenario 
from the usual inverse seesaw extended SM.   
\end{abstract}
%
%
\maketitle
\section{Introduction}
\label{sec:intro}
ATLAS and CMS collaborations have already made huge impacts to unravel any possible signatures of physics 
beyond the Standard Model (BSM) at the LHC \cite{lhc1,lhc2}. Although no signature of new particles has yet 
been found by the direct search, there are many reasons for pursuing our search in BSM physics at the LHC. 
Undoubtedly, one of the most compelling motivations comes from the existence of non-zero neutrino masses 
and substantial mixing among the three light neutrino states (see for example, \cite{GonzalezGarcia:2007ib,
Gonzalez-Garcia:2015qrr}). 
In this frontier, the neutrino oscillation experiments have made significant progress to measure the
mass square difference and the mixing angles of the neutrinos with unprecedented precision, which indicates that at least two of the light neutrino states 
have to have tiny non-zero masses. As a natural consequence, many proposals have been put forward over the years to 
accommodate the neutrino masses and mixing in the theory of which various {\it Seesaw Mechanisms} have remained of 
primary interest \cite{Minkowski:1977sc,Mohapatra:1979ia,
GellMann:1980vs,Yanagida:1979as,Glashow:1979nm,Schechter:1981cv,Schechter:1980gr,Weinberg:1979sa,Weinberg:1980bf,Magg:1980ut,Cheng:1980qt,Foot:1988aq}.

The simplest seesaw extension, known as the {\it Type-I Seesaw mechanism} \cite{Minkowski:1977sc,Mohapatra:1979ia,
Yanagida:1979as,GellMann:1980vs} is accompanied with additional 
Majorana right-handed neutrinos (${\nu}_{Ri}$). Light neutrinos gain tiny non-zero masses 
by virtue of their mixings with these heavy neutrino states. In such scenarios, the mass scale of these Majorana neutrinos 
has to lie very close to gauge coupling unification scale $M_G \sim 10^{16}$~GeV in order to account for the tiny 
neutrino masses. Such a massive ${\nu}_R$ completely decouples from the low energy theory and remains out of the kinematic 
reach of the LHC. One can of course bring down this mass scale to TeV range, but at the cost of a very small Dirac 
neutrino Yukawa coupling ($\sim 10^{-6}$) which again makes any possible phenomenological aspects of the heavy neutrinos 
unforeseeable at the LHC\footnote{Such small couplings can only be probed with very high precision colliders and naturally 
lepton colliders are better suited to serve the purpose \cite{Antusch:2015mia}.}. Thus the lack of predictability of this simplistic scenario 
has forced theorists as well as experimentalists to study models which can readily be probed at the LHC with the existing 
data. One such scenario is called the {\it inverse seesaw}
mechanism \cite{Mohapatra:1986aw,Nandi:1985uh,Mohapatra:1986bd} where the SM particle content is augmented by two singlet neutrinos 
with opposite lepton numbers 
($+1$ and $-1$). The generic form of the light neutrino mass can be expressed as  $m_\nu \sim (m_D^2/M_R^2) \mu$, where $m_D \sim Y_\nu v$ 
represents the Dirac neutrino mass term, $v$, the electroweak VEV, $Y_\nu$, a generic Yukawa coupling and $\mu$, a lepton number 
violating ($\Delta L=2$) mass parameter which can be kept naturally small \cite{tHooft:1979rat}. The presence of this small ($\sim $ eV) 
$\Delta L=2$ mass term helps to keep the sterile neutrino mass scale $M_R$ close to TeV, i.e., within the reach of LHC, 
with order one Yukawa coupling. This feature of the model leads to a plethora of testable phenomenological consequences \cite{Cao:2017cjf,Karmakar:2016cvb,Sinha:2015ooa,Arganda:2015naa,Abada:2012cq,Abada:2011hm,Abada:2010ym,Arganda:2014dta,Arganda:2015ija,
DeRomeri:2016gum,Arganda:2017vdb} which have 
been studied quite extensively in the context of present and future collider experiments \cite{Chen:2011hc, Das:2012ze,Bandyopadhyay:2012px, Dev:2013wba,Das:2014jxa,Deppisch:2015qwa,Das:2015toa,Mondal:2016kof, Banerjee:2015gca, Das:2017pvt}. For the RH neutrino production through vector boson fusion and higher order corrections, see \cite{Alva:2014gxa,Degrande:2016aje}.
Note that, although the 
  heavy neutrino states now lie within the kinematic reach of the LHC,
the production processes are still driven by the charged/neutral current weak interactions, through 
active-sterile neutrino mixing, that depend crucially on the mass scale of the RH neutrinos $M_R$. Hence,  
the cross-section falls rapidly with increasing heavy neutrino masses and with smaller mixings. 
The production cross-section of the RH neutrinos can be drastically enhanced in presence of leptoquarks (LQ). Leptoquarks, being charged under $SU(3)_C$, are copiously produced at the LHC. The right-handed neutrinos can result from the decays of LQs. It is important to note that in this case, the RH neutrino production does not depend on the active-sterile mixing. Therefore, RH neutrinos can be probed at the LHC irrespective of their mixing with the active component as long as they decay inside the detector.

Introduction of LQ (for a recent review see \cite{Dorsner:2016wpm}) to the SM Lagrangian is motivated from a 
quite different viewpoint. In fact, 
in the Pati-Salam model that LQs are a natural outcome of unification of quarks and leptons \cite{Pati:1974yy}. 
The presence of these new exotic particles has been further motivated by the simple grand unified gauge groups of $SU(5)$ 
\cite{Georgi:1974sy} and $SO(10)$ \cite{Georgi:1974my,Fritzsch:1974nn}. While both vector (spin-one) and scalar LQ (spin-zero) 
states are possible in local quantum field theories, the scalar states are more attractive as they do not lead to any potential
problems related to loop corrections \cite{Altmannshofer:2016oaq,Becirevic:2016yqi}. Moreover, it has recently been shown that the scalar LQs are indeed very 
useful to explain various $B$-physics anomalies like $R_K$ \cite{Becirevic:2015asa} or $R_{{D}^*}$ \cite{Dorsner:2013tla}. Explanation of both $R_K$ and $R_{{D}}$ anomalies with a scalar LQ with the hypercharge $Y=\frac{1}{6}$ 
would be possible if one includes a new interaction between a scalar LQ and a right handed neutrino \cite{Becirevic:2016yqi}. Keeping in mind that the presence
of a LQ-$\nu_R$ coupling is highly motivated to accommodate B-physics anomalies, in this work we will go one step further. 
We will introduce additional sterile neutrinos in the model to comply with all existing experimental data on light neutrino 
mass and mixing angles while the mass of the heavy neutrino states have been kept smaller than the leptoquark states. In practice, 
we incorporate {\it inverse seesaw} to the SM extended by leptoquarks for our study. We show that the heavy neutrinos can be copiously generated from the decays of the scalar LQs which can be produced via 
strong interactions thanks to their $SU(3)$ interactions. The relevant couplings and masses of the LQs and the 
heavy neutrinos are chosen in a way such that they are consistent with the experimental constraints and at the same time 
maximize the heavy neutrino productions from the cascade decay. Moreover our collider study shows that such a scenario 
has the potential of probing heavy neutrino masses up to a much higher range compared to that in the usual neutrino mass models. 

The paper is organized as follows. In Sec.~\ref{sec:mod}, we introduce the model. Following that, in Section.~\ref{sec:stsconstbps}, we discuss the present experimental constraints on the RH neutrinos and the LQs. The collider signatures 
have been discussed in Section.~\ref{sec:collans}. Finally, we present our 
conclusion.  
\section{Inverse Seesaw Mechanism in the SM extended by Leptoquarks \label{sec:mod}}
The Lagrangian which we will consider in this work can be cast as
\bea
{\mathcal L} = \cl_{SM}+\cl_{\Delta}+\cl_{IS}~.
\label{eq:lag}
\eea
$\cl_{\Delta}$ includes the interaction terms between the leptoquarks and the SM particles while $\cl_{IS}$ 
comprises of the relevant terms for generation of the neutrino mass and mixing angles. The association between $\Delta$ and $\cl_{IS}$ have been realized through right handed 
neutrino states $\nu_{Ri}$ whose mass can vary $\mathcal {O}(1) \rm ~eV \le M_{Ri} \le \mathcal {O}(1)$~TeV.
Moreover the inverse seesaw extension of the SM requires three new fermionic singlet fields ${X}_i$ ($i=1,2,3$)
with lepton numbers $+1$ in contrast to the $\nu_{Ri}$ states of lepton numbers $-1$ respectively\footnote{Note that in the minimal inverse seesaw extension of the SM, two pairs of the singlet field would be sufficient 
to satisfy all neutrino data \cite{Abada:2014vea}.}. 
\bea
{\mathcal L}& \in & \varepsilon_{ab} Y^{ij}_\nu 
{\nu}^R_i {L}^a_j {H}^b 
+M_{R_i}{\nu}_{Ri}{X}_i+
\frac{1}{2}\mu_{X_{ij}}{X}_i{X}_j  ~,
\label{eq:lag}
\eea
\noindent where $i,j = 1,2,3$ denote generation indices.
In the above $L_i$ denotes three generation SU(2)
 lepton doublets.
$M_{R_i}$ represents the right-handed neutrino
bilinear term which conserves lepton number. 
The total lepton number $L$ is no longer a good
quantum number because of non-vanishing $\mu_{X}$, though 
$(-1)^L$ is still a good symmetry.  
Clearly, as $\mu_{X_i}\rightarrow 0$, lepton number conservation is restored,
since $M_R$ does not violate lepton number. 

The Leptoquark part of the Lagrangian for a scalar LQ state 
$\Delta$ transforms as $\Delta \in \left(3,2,\frac{1}{6}\right)$ under the SM gauge group can be written as \cite{Dorsner:2016wpm,Becirevic:2016yqi}
\bea
\cl_{\Delta} &=& {\overline d_R Y_L (\widetilde \Delta)^\dagger
  L +  \overline Q Y_R \Delta \nu_R +  \mathrm{ h.c.}}
\label{eq:lq1};
\eea
where $\widetilde \Delta = i\sigma_2 \Delta^*$. Defining mass eigenstates for 
a fermionic field $\psi^i_{L,R}$ as $\psi^{mi}_{L,R}=U^{i}_{L,R} \psi^i_{L,R}$
\cite{Becirevic:2016yqi}, one can get (for simplicity we remove the superscript $m$ from the fields in mass basis.)
\begin{align}\label{eq:lq2}
&\cl_{\Delta} =Y_L^{ij} \overline d_R^i U_{\rm PMNS}^{jk} \nu_L^k\Delta^{(-1/3)} - Y_L^{ij} \overline d_R^i \ell_L^j\Delta^{(2/3)}  \cr
& + Y_R^{ij} \overline u_L^i \nu_R^j\Delta^{(2/3)} + Y_R^{ij} \overline d_L^{k} V_{\rm CKM}^{ki} \nu_R^j\Delta^{(-1/3)} + \mathrm{ h.c.}, 
\end{align}
where $Y_{L} \to {U_{R}^{d\dag} Y_{L} U_L^{\ell}}$ and
$Y_{R} \to {U_{L}^{u\dag} Y_{R} U_R^{\nu}}$ have been used. Similarly,
$U_{\rm PMNS}= {U_L^\ell  U_L^{\nu \dag}}$, and $V_{\rm CKM} = U_L^{d}U_L^{u\dag}$ 
represent the Pontecorvo-Maki-Nakagawa-Sakata and the Cabibbo-
Kobayashi-Maskawa matrices. Now assuming the Yukawa couplings to the first 
generations of the quarks/leptons as zero to become consistent with the atomic parity 
violation experiments~\cite{Dorsner:2016wpm}, we represent the matrices as follows \cite{Becirevic:2016yqi}: 
\[
Y_{L,R} = \left( \begin{matrix}
  0 & 0 & 0\\
  0 & Y_{L,R}^{22} & Y_{L,R}^{23}\\
  0 & Y_{L,R}^{32} & Y_{L,R}^{33}
\end{matrix}\right).
\] 

Here we primarily consider productions of $\Delta^{(2/3)}$ states  because of its direct coupling with $t$ quarks and the third generation RH neutrinos $\nu_{R_3}$.
In the following we consider two scenarios. (i) Yukawa couplings that can 
potentially explain the $B$-physics anomalies. Here $Y_{L,R}^{2i}$ can be small, while $Y_{L,R}^{3i}$ can be large. In particular there is a lower bound on
  $Y_{L}^{33} \geq 1.5$ \cite{pc:olcyr} which we will use for our
  calculation (ii)
Yukawa couplings of the 2nd and 3rd generations are unconstrained from 
$B$-physics, though the choice of $Y_{R}^{33}$ is completely driven by the
  fact that we need to maximize the branching ratio LQ $\rightarrow t N_{\tau}$.
\subsection{Neutrino masses in Inverse seesaw}
\label{subsec:numass}
We consider a general framework with
three generations for the sterile neutrinos, namely $\nu_{Ri}$ and $X_i$. 
Consequently, one has the following symmetric $(9\times9)$
mass matrix $\mathcal{M}$ in the basis
$\{\nu,{\nu_R},X\}$, 
\begin{eqnarray}
{\cal M}&=&\left(
\begin{array}{ccc}
0 & m^{T}_D & 0 \\
m_D & 0 & M_R \\
0 & M^{T}_R & \mu_X \\
\end{array}\right) \ ,
\label{nmssm-matrix}
\end{eqnarray}
Where, $m_D= \frac{1}{\sqrt 2} Y_\nu v$ and $M_R$, $\mu_X$ are  
$(3\times3)$ matrices in family space.
Assuming $(m_D,\mu_X \ll M_R)$ the diagonalisation results in an
effective Majorana mass matrix for the light
neutrinos
\begin{equation}
\label{eqn:nu}
    m_\nu = {m_D^T M_R^{T}}^{-1} \mu_X M_R^{-1} m_D
          = \frac{v^2}{2} Y^T_\nu (M^T_R)^{-1} \mu_X M_R^{-1} Y_\nu.
\end{equation}
The most important aspect of the inverse seesaw mechanism is that the smallness of the light neutrino masses is
directly controlled by the scale of $\mu_X$. Having this small dimensionful term in the Lagrangian is technically
natural in the sense of 't Hooft \cite{tHooft:1979rat}, as in the limit of
vanishing $\mu_X$ one recovers the lepton number symmetry. 
The lepton number
conserving mass parameters $m_D$ and $M_R$ can easily accommodate
large (natural) Yukawa couplings ($Y_\nu\sim {\mathcal{O}}(1) $) and a
right-handed neutrino mass scale around the TeV, see Eq.~(\ref{eqn:nu}).
In analogy to a type-I seesaw, one can define an effective right-handed
neutrino mass term $M$ such that
\begin{equation} 
\label{M-def}
M^{-1} = (M^T_R)^{-1} \cdot \mu_X \cdot M_R^{-1}.
\end{equation}
With this definition,  the light neutrino mass matrix can be cast in a way which strongly resembles a standard (type-I) seesaw equation
\begin{equation} \label{M-def2}
 m_\nu = \frac{v^2}{2} Y^T_\nu M^{-1}  Y_\nu.
\end{equation}
This effective light neutrino mass matrix $(m_\nu)$ can be diagonalized as
\begin{equation}
    U_{PMNS}^T m_\nu U_{PMNS} = \textrm{diag}\;m_i\,.
\label{eqn:dia}
\end{equation}
In
the subsequent analysis, we assume $M_R$ and $Y_\nu$ are diagonal 
($M_{R_{ij}} = \textrm{diag}\;M_{R_{ii}}, Y_{\nu_{ij}} = \textrm{diag}\;Y_{\nu_{ii}}$).
This choice of $\textrm{diag}\;Y_{\nu_{ii}}$ 
naturally ameliorates constraints 
from lepton flavor changing  processes. However, $\mu_{Xij}$ 
is not flavor diagonal and its structure can be determined by using Eq.~\ref{M-def}, Eq.~\ref{M-def2}, and Eq.~\ref{eqn:dia}  
where  the present results on neutrino data \cite{Gonzalez-Garcia:2015qrr} have been used. 

In our analysis, we have  used the best-fit values of the mass 
square differences and the mixing angles to fit $\mu_X$. The light neutrino masses and the PMNS mixing matrix have been depicted in Tab.~\ref{tab:osc}.
\begin{table}[h!]
\begin{center}
\begin{tabular}{||c|c||} 
\hline 
Parameters & Values   \\
\hline\hline  
${\rm sin}^2 \theta_{12}$ & $0.304^{+0.013}_{-0.012}$ \\
\hline
${\rm sin}^2 \theta_{23}$ & $0.452^{+0.052}_{-0.028}$ \\
\hline
${\rm sin}^2 \theta_{13}$ & $0.0218^{+0.001}_{-0.001}$ \\
\hline
$\Delta m^2_{21}~{\rm eV}^2$  & $(7.50^{+0.19}_{-0.17})\times 10^{-5}$ \\
\hline
$\Delta m^2_{32}~{\rm eV}^2$ & $(2.457^{+0.047}_{-0.047})\times 10^{-3}$ \\
\hline
PMNS  \cite{Gonzalez-Garcia:2015qrr} & {\tiny $\left(\begin{array}{ccc}
0.801 - 0.845 & 0.225 - 0.517  &0.246 - 0.529 \\
0.514 - 0.580 & 0.441 - 0.699 &0.464 - 0.713  \\
0.137 - 0.158 & 0.614 - 0.793 & 0.590 - 0.776\\
\end{array}\right)$}\\ 
\hline\hline 
\end{tabular}
\caption{Three flavor neutrino oscillation data obtained from global fit for normal hierarchy in neutrino masses 
and the PMNS matrix elements with their $3\sigma$ allowed ranges. We have only used the best-fit values of the mass 
square differences and the mixing angles to fit an off-diagonal $\mu_X$ in order to fit the neutrino oscillation data.}
\label{tab:osc} 
\end{center}
\end{table} 
\subsection{Production and decay of Leptoquarks}
\label{subsec:prod_dec_lq}
\begin{figure}
\begin{center}
\includegraphics[width=10.0cm]{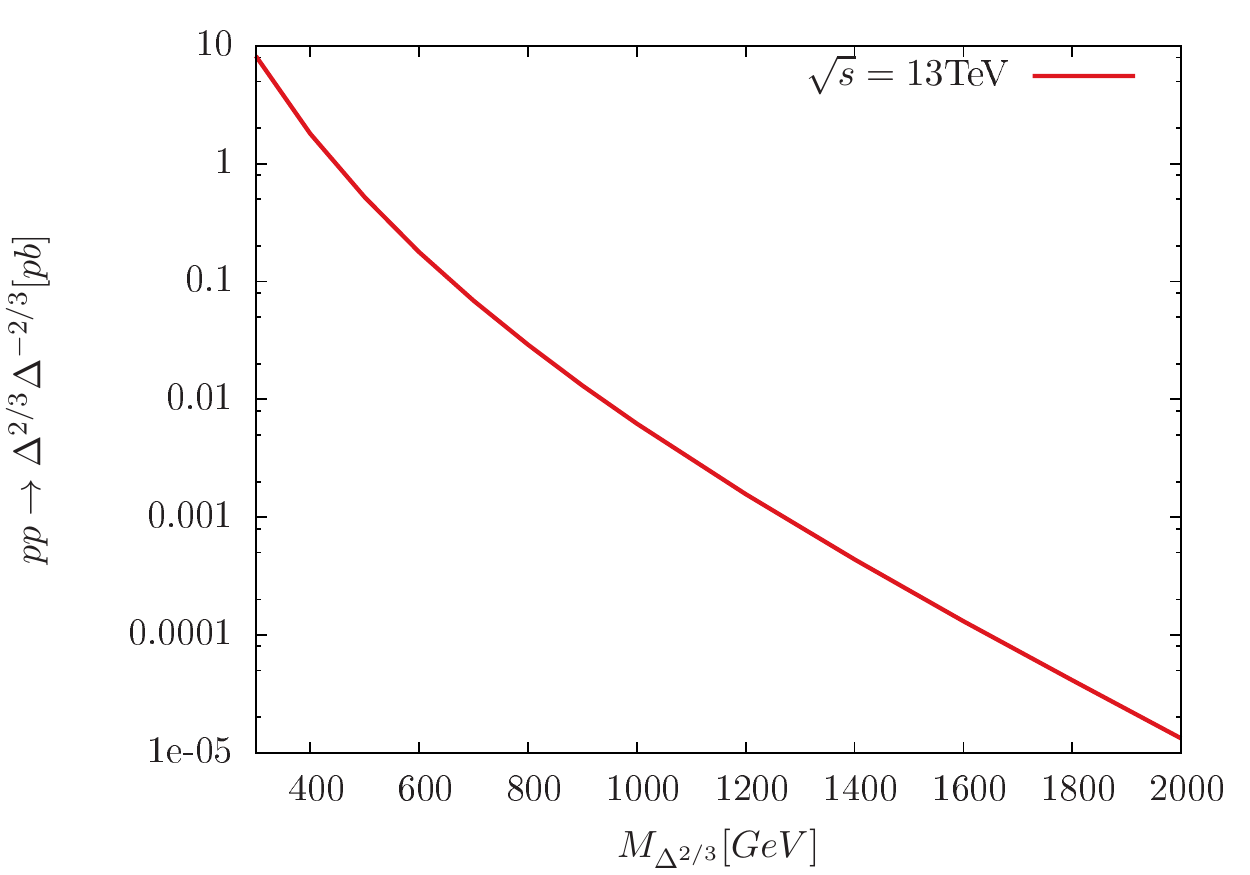}
\caption{Pair production cross-section of leptoquarks at the LHC with 13 TeV center-of-mass energy as a function of the leptoquark mass. }
\label{fig:lqa1}
\end{center}
\end{figure}

Being triplet under $SU(3)_C$, leptoquarks have gauge couplings with gluons. Therefore, at the LHC, leptoquarks are produced 
in pairs via gluon-gluon and quark-antiquark initiated processes. Gluon-gluon initiated processes primarily proceed through a leptoquark 
exchange in the $t(u)$-channel or a gluon exchange in the $s$-channel. On the other hand, quark-antiquark initiated processes 
take place only via a gluon exchange in the $s$-channel and hence, suppressed compared to the $t(u)$-channel leptoquark exchange. 
Leptoquark can also couple to a quark-lepton pair via Yukawa interactions as shown in Eq.~\ref{eq:lq2}. Therefore, quark-antiquark initiated 
processes may also proceed through a lepton (charged or neutral) exchange in the $t(u)$-channel. However, the Yukawa couplings of the 
leptoquarks with the first generation of quarks and leptons are severely constrained from the atomic parity violation experiments 
\cite{Dorsner:2016wpm}, making this contribution small. In order to generate the model files for further simulation, we have incorporated the new physics Lagrangian in 
FeynRules (v2.3.13) \cite{Christensen:2008py,Alloul:2013bka} and subsequently generated model files are used to compute particle spectrum, 
relevant branching ratios and cross-sections via MadGraph5 (v2.4.3) \cite{Alwall:2011uj,
Alwall:2014hca} with NNPDF23lo1 \cite{Ball:2012cx,Ball:2014uwa} parton distribution function (PDF). We have used dynamic 
definition \cite{madgraph_scale} of the factorization scale for 
the PDF and renormalization scale for evaluating the QCD coupling. Being mainly QCD driven, the pair production of leptoquarks \cite{Diaz:2017lit} gets 
significant corrections from the higher order processes. In Ref.~\cite{Mandal:2015lca}, next-to-leading (NLO) order in QCD $k$-factor for leptoquark 
pair production at the 13 TeV LHC is estimated to be $\sim~1.4$. In our calculation, we have included the $k$-factor.
The QCD pair production cross-section of the leptoquarks ($\sigma(pp\to\Delta^{+2/3}\Delta^{-2/3})$) 
as a function of the leptoquark mass ($M_{\Delta^{\pm2/3}}$) at the LHC with 13 TeV center-of-mass energy is presented in 
Fig.~\ref{fig:lqa1}. Being strongly interacting, leptoquark pair production cross-sections are rather large at the LHC as it can be 
seen from Fig.~\ref{fig:lqa1}. The pair production cross-section varies from few picobarn to few femtobarn for the LQ mass between few hundred GeV to TeV. 

After being produced, leptoquarks decay into a quark-lepton pairs via the Yukawa interactions of Eq.~\ref{eq:lq2}. 
For example, $\Delta^{\pm2/3}$ can decay into a down type quark and charged lepton pair or a up type quark and 
right-handed neutrino pair. The decay of  $\Delta^{\pm2/3}$ into a down type quark and charged lepton pair is 
proportional to $Y_L$ whereas, the decay into a up type quark and right-handed neutrino pair is proportional 
to $Y_R$. For simplicity, we have assumed that leptoquark Yukawa matrices are diagonal and leptoquarks can dominantly 
couple to the third generation of SM fermions\footnote{ This choice is 
motivated as leptoquark decaying into a top quark and heavy neutrino 
results in higher jet, lepton multiplicities and larger missing energy in the final state.}.
Similarly, as mentioned earlier, we have also chosen the neutrino Yukawa matrix ($Y_\nu$) 
to be diagonal in order to avoid constraints arising from non-observation of lepton flavor 
violating decays. As a result of this choice the third generation RH neutrino (which will be denoted as $N_\tau$ in the mass basis) can only directly decay into 
a tau and W-boson or a light $\nu_\tau$ and Z (with smaller branching ratio) through its left handed components which compells us to look for tau-enriched final states. This will also help us to reduce SM backgrounds, atleast for the signal reagions we will be dealing with. We note in passing that keeping
the flavor diagonal structure for $Y_{L,R}$, one may obtain $\mu$ rich final states for larger values of $Y^{22}_{R}$ ($Y^{22}_{R} >> Y^{22}_{L}$ and
$Y^{22}_{L,R} >>Y^{33}_{L,R}$), but then one may face relatively difficult task in regard to
$c$-jet tagging.

\begin{figure}
\begin{center}
\includegraphics[width=7.5cm,height=7.0cm]{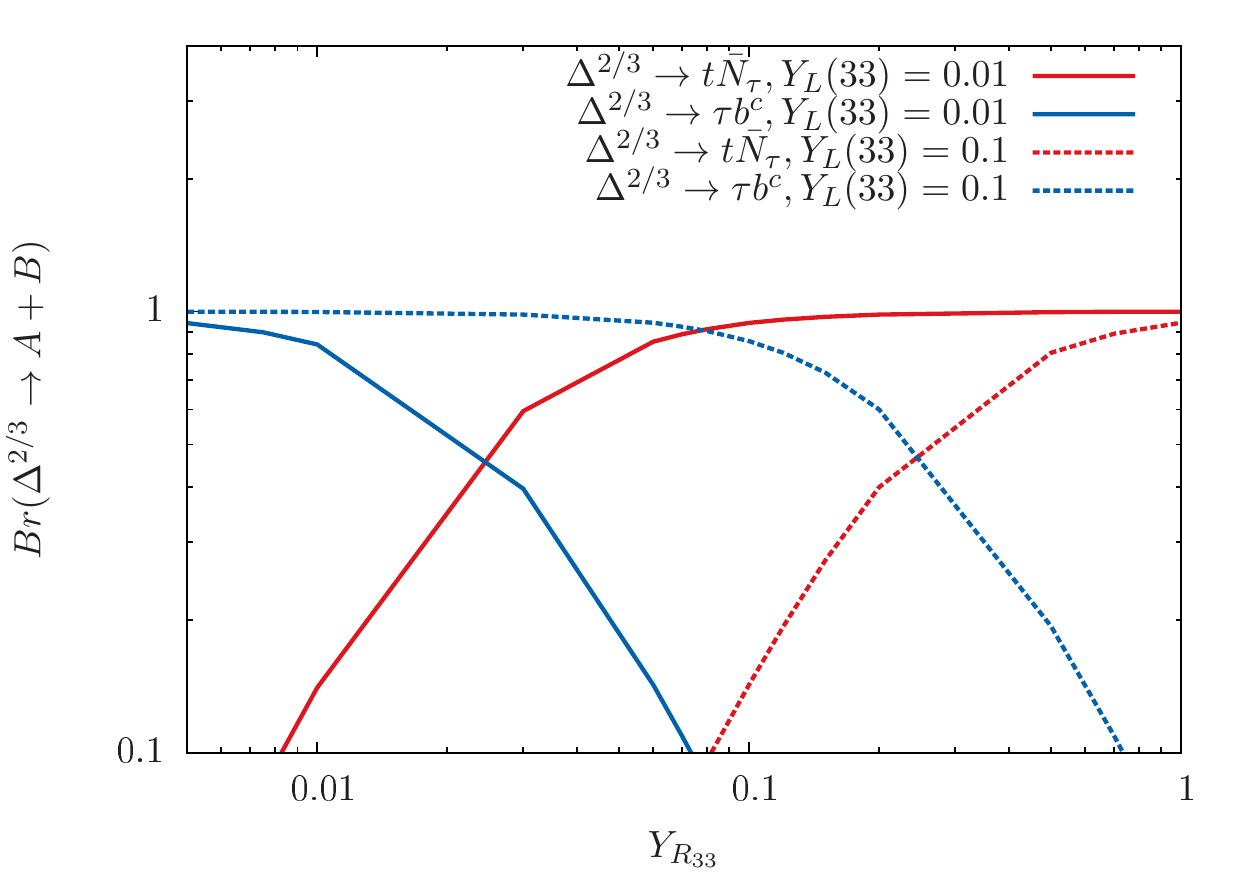}
\includegraphics[width=7.50cm,height=7.0cm]{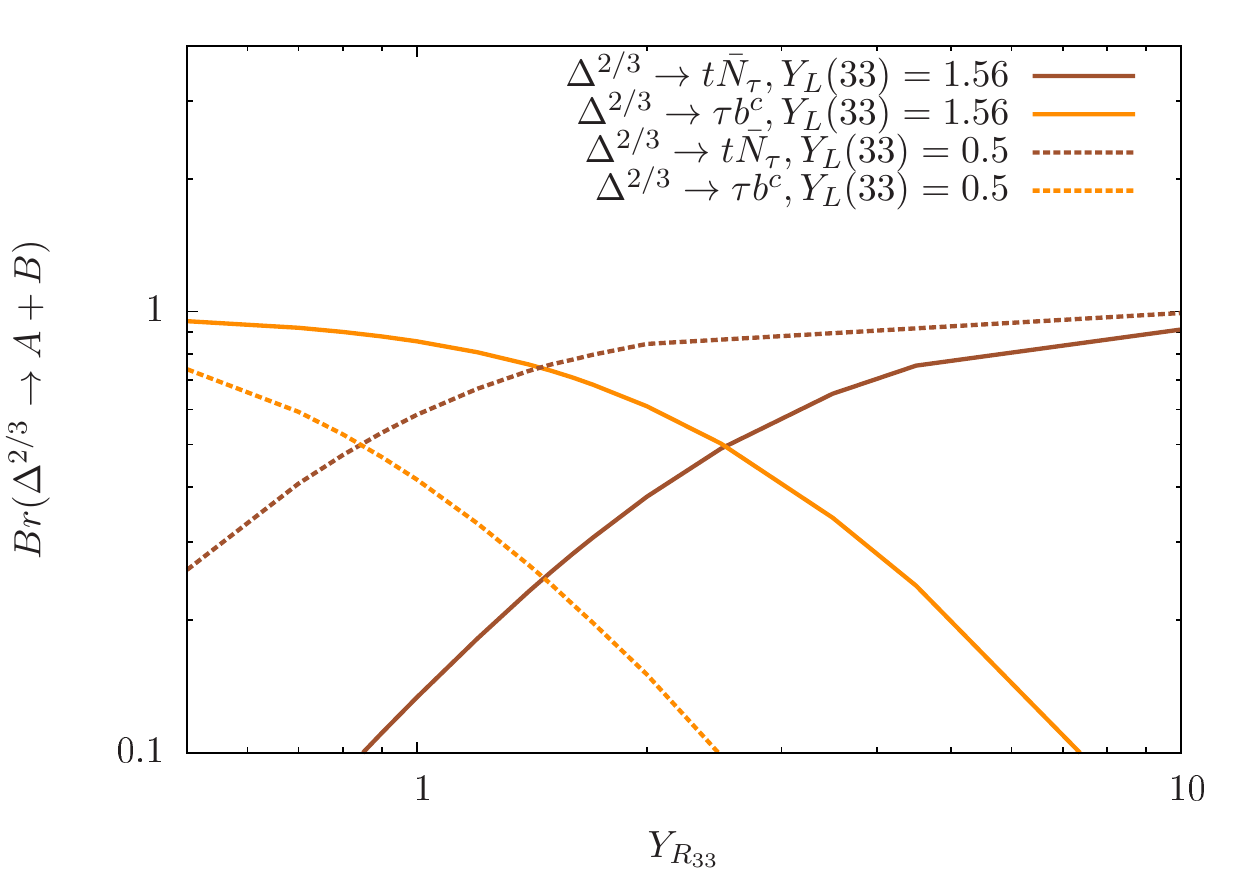}
\caption{Branching ratios of $\Delta^{\pm2/3}$ into $t N_\tau$ and $b \tau $ pairs as a function of $Y_R^{33}$ for different values of $Y_L^{33}$. {Left and right panel correspond to  850 GeV and 1 TeV leptoquark masses, as illustrative examples.}}
\label{fig:lqb1}
\end{center}
\end{figure}

Thus, in this framework, $\Delta^{+2/3}$ dominantly 
decays into a top and third generation right-handed neutrino pair ($t N_\tau$) or a bottom quark and tau lepton pair 
($b \bar \tau $) depending on the relative values of $Y_R^{33}$ and $Y_L^{33}$, respectively. In Fig.~\ref{fig:lqb1}, 
we have presented the branching ratio of $\Delta^{\pm2/3}$ into $t N_\tau$ and $b \tau $ pairs as a function of $Y_R^{33}$ 
for different values of $Y_L^{33}$, and write down the analytic expression of the decay widths in the appendix.

\begin{figure}
\begin{center}
\includegraphics[width=8.50cm,height=7.0cm]{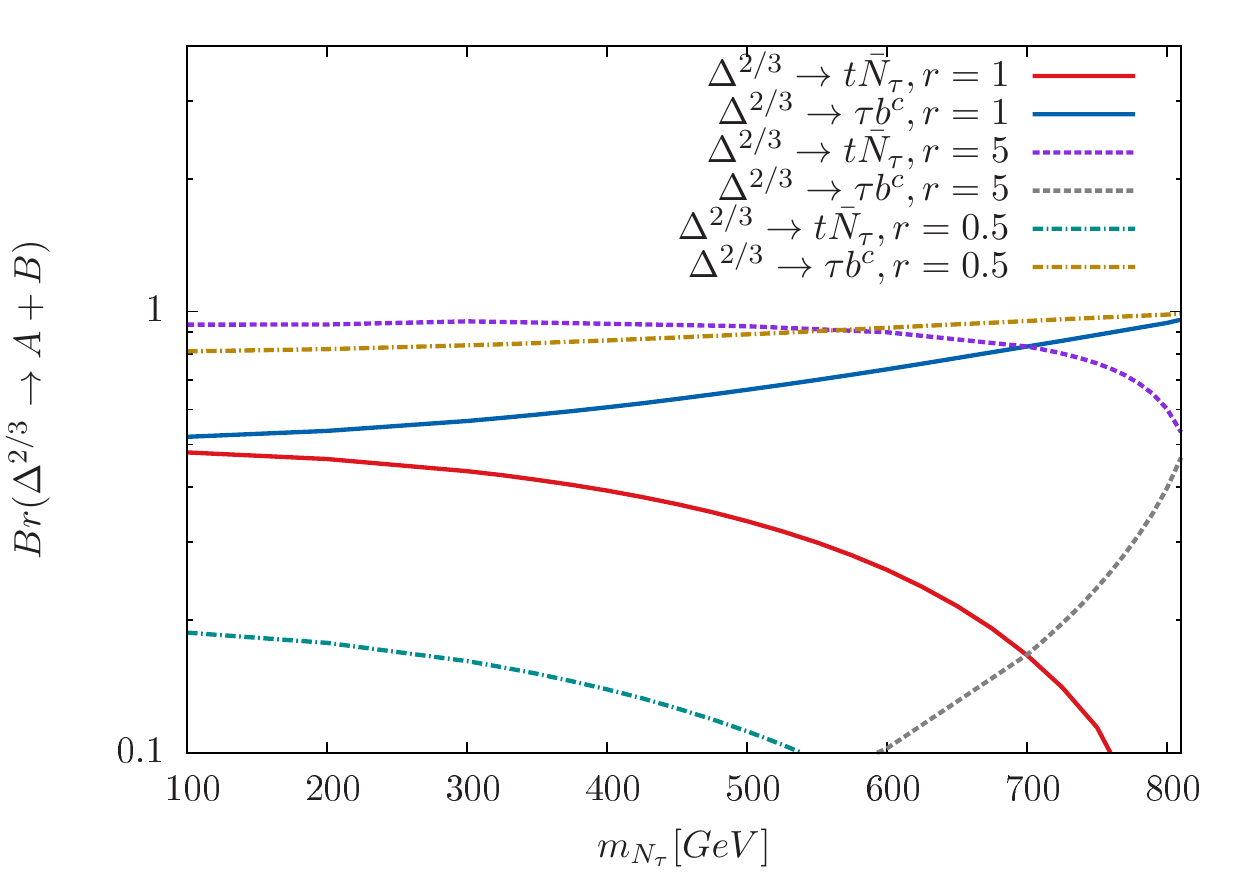}
\caption{Variation of branching ratios of $\Delta^{\pm2/3}$ into $t N_\tau$ and $b \tau $ pairs as a function of heavy neutrino mass $m_{N_\tau}$. The LQ mass is set to 1 TeV and $r=\frac{Y_{R_{33}}}{Y_{L_{33}}}$. }
\label{fig:lqb2}
\end{center}
\end{figure}

Fig.~\ref{fig:lqb1} clearly shows that $tN_\tau$ decay mode dominates for larger values 
of $Y_R^{33}$ compared to $Y_L^{33}$. The off-shell decays 
for example $\Delta^{2/3} \to {\Delta^{1/3}} W^*$ does not occur 
in this scenario due to mass degeneracy of the two LQs.  In order to enhance the production of right handed neutrinos from the decay of 
$\Delta^{\pm2/3}$, smaller $Y_L^{33}$ compared to $Y_{R}^{33}$ is preferred. Note that, for larger $Y_L^{33}$, the bounds 
on $\Delta^{\pm2/3}$ mass are much stronger from resonance searches in the lepton-jet invariant mass distribution at the LHC \cite{CMS:2016hsa}. 
Therefore, in our analysis, we consider $Y_R^{33}>Y_{L}^{33}$ hence, $\Delta^{\pm2/3}$ dominantly decays into $t N_\tau$ 
pairs with almost 100\% branching ratios.

In addition to the variation w.r.t Yukawa, we also show the variation w.r.t the RH neutrino mass. In Fig.~\ref{fig:lqb2}, we show the variation of the branching ratios of $\Delta^{2/3}$ with the RH neutrino mass $m_{N_\tau}$. The LQ mass is set to 1 TeV and we define $r$ as the ratio of
    $\frac{Y_{R_{33}}}{Y_{L_{33}}}$ to delineate the dependence of the branching fraction on the $m_{N_\tau}$. Clearly,
$Br(\Delta^{2/3} \rightarrow tN_\tau)$ would be the most dominant channel for $r>1$. For representation purpose we assume $r=5$, and we observe that the $ Br(\Delta^{2/3} \longrightarrow tN_\tau)$ is always 100\% if
phase space is allowed. The fall in $Br(\Delta^{2/3} \rightarrow tN_\tau)$ occurs near the kinematic threashold $m_{N_\tau} \sim 827$ GeV for the LQ mass 1 TeV. After the kinematic threshold, additional three body modes such as, 
$\Delta^{2/3} \rightarrow t W l, t Z \nu $ with branching ratio $30-40\%$ also  open up. In our subsequent analysis, we work with higher $Y_R^{33}$, and relatively smaller mass of $N_{\tau} $ as compared to the LQ, so that the two body decay mode $\Delta^{\pm2/3} \to t N_{\tau}$ is almost 100$\%$.

The produced  $N_\tau$ subsequently decays into a $\tau W$ pair or a $\nu_\tau Z$ \footnote{Though the decay 
widths of $N_\tau$ into $\tau W$ or $\nu_\tau Z$ are suppressed by the small left-handed and right-handed neutrino mixing angle, 
but in absence of any other decay modes, $N_\tau$ decays into $\tau W$ and $\nu_\tau Z$ pair with 67\% and 33\% branching 
ratios, respectively.} pair. Therefore, the pair production of $\Delta^{\pm2/3}$ at the LHC gives rise to multiple top-quarks, 
tau-leptons, $W$ and/or $Z$-bosons in the final state. Subsequent decays of those heavier SM particles result into multiple 
leptons, jets and missing transverse energy ($E_T\!\!\!\!\!/~$) signature at the LHC. Before going into the details of signal 
and background analysis, let us have a brief discussion on the present status of the heavy neutrino and leptoquark searches in 
collider and other experiments which lead to our choices of benchmark points for subsequent collider study.
\section{Present status, constraints and choice of Benchmark Points 
\label{sec:stsconstbps} }
The collider signatures of right-handed neutrinos have been extensively studied in the literature 
\cite{Keung:1983uu,Datta:1993nm,Almeida:2000pz,Panella:2001wq,Han:2006ip,delAguila:2007qnc,
Huitu:2008gf,Atre:2009rg,Keung:1983uu,Datta:1993nm,Almeida:2000pz,Panella:2001wq,Han:2006ip,Chen:2011hc,Alva:2014gxa,Degrande:2016aje,
Das:2016hof,Ruiz:2017yyf}.  
Usually, the search for a heavy Majorana neutrino is driven by lepton number violating final states which 
provide a smoking-gun signals for such scenarios. However, within the framework of inverse seesaw, the lepton 
number is broken by the $\mu$-parameter which is extremely small and therefore, the usual same-sign dilepton signal 
is expected to be much weaker in this scenario compared to a trilepton signal \cite{Das:2015toa}. 
In fact, the smallness of 
$\mu$ results in suppression of any lepton number violating (LNV) processes in this framework. Thus in the considered scenario, the first-generation 
Dirac neutrino Yukawa coupling $Y^{11}_{\nu}$, which is highly constrained for Type-I seesaw mechanism from non-observation of 
neutrino-less double beta decay ($0\nu\beta\beta$) \cite{Deppisch:2015qwa}, remains practically unconstrained. 
The experimental collaborations have put constraints on the heavy Majorana neutrino masses depending on its choice of 
Dirac neutrino Yukawa couplings from the search of same flavor opposite-sign dilepton final state \cite{Abulencia:2007rd,
Chatrchyan:2012fla,ATLAS:2012yoa,Khachatryan:2015gha,Aad:2015xaa,Khachatryan:2016olu}. In the inverse 
seesaw scenario, these constraints are trivially satisfied. Comparatively, some other final states with various lepton-jet 
multiplicities can provide a more stringent constraint on the inverse seesaw scenario that can probe a mixing angle 
up to $\sim 10^{-2}$ for a heavy neutrino mass around 200 GeV at 14 TeV run of the LHC with a luminosity around 
1000 ${\rm fb}^{-1}$ \cite{Das:2015toa}. In this work, we aim to probe up to a much higher RH neutrino mass range. This is achievable, as our RH neutrino production is not limited by the active-sterile mixing which is further constrained from precision studies \cite{delAguila:2008pw,Akhmedov:2013hec,Basso:2013jka,Antusch:2015mia,Antusch:2014woa}. Instead, the heavy neutrino is produced from leptoquark decays, that has strong interaction. This helps to obtain a relatively larger RH neutrino production cross-section. 
Another potential source of the constraints on the 
neutrino sector may arise from the lepton flavor violating (LFV) decays   that arise due to non-diagonal $Y_{\nu}$ 
or $M_R$. In this analysis, we consider  these two matrices strictly diagonal which leads to vanishing contribution of the new physics 
contribution to the LFV decays.  

Experimental collaborations have also searched for any possible hints for scalar leptoquarks at the LHC \cite{Aaboud:2016qeg,
Sirunyan:2017yrk}. The favored final state to look for these exotic particles has been $\ell\ell jj$, where, 
$\ell=e, \mu$, $\tau$. The non-observation of any new physics events have ruled out LQ up to 
masses as large as 1100 GeV. However, these existing limits are subjected to a particular choice of $Y_L$ 
that forces the leptoquarks to decay dominantly into a charged lepton and a light (or bottom) quark. In this work, however, we 
intend to explore a different coupling of the LQ, namely, $Y_R$ which forces the LQ to decay via the heavy neutrinos. 
Therefore, taking a clue from the obtained branching ratio distribution of the LQs as shown in Fig.~\ref{fig:lqb1}, 
we keep  $Y_L$ smaller in comparison to $Y_R$ for all our benchmark points so that relatively lighter LQ masses can be probed which are still allowed by the LHC. 
However, the leptoquark decaying via heavy neutrinos can give rise 
to the same kind of multi lepton signals which has usually been studied in
the context of supersymmetric searches \cite{Aad:2014pda,Aaboud:2017dmy,Aad:2016tuk}. So far all such search results have turned out to be consistent 
with the SM predictions and hence these results can constrain the LQ as well as heavy neutrino masses in our scenario.

As has been mentioned, in this work, we have considered the pair production of $\Delta^{\pm2/3}$ and its subsequent decay 
into a top and a heavy neutrino. Hence, the collider phenomenological aspects of our study are 
determined by five parameters namely, leptoquark mass ($m_{\Delta^{\pm2/3}}$), its couplings 
$Y_{L,R}^{33}$, right-handed neutrino mass ($m_{N_\tau}$) and its coupling $m_D^{33}$. The 
pair-production cross-section $\sigma(\Delta^{\pm2/3}\Delta^{\mp2/3})$ is determined by the 
choice of $m_{\Delta^{\pm2/3}}$ and its couplings. While, relative strength of $Y_{R}^{33}$ 
and $Y_L^{33}$ determines the decay of $\Delta^{\pm2/3}$ into right-handed neutrino ($N_\tau$),  
the mass splitting between $\Delta^{\pm2/3}$ - $N_\tau$ determines the shape of characteristic 
signal distributions. In order to present the numerical results of our analysis, we have 
chosen three benchmark points (BP). The relevant parameters for the collider phenomenology for 
those three BPs are listed in Tables~\ref{tab:param} and \ref{tab:bp}. We also present a bench-mark point 
(BP2 in Table~\ref{tab:bp}) which can potentially explain $B-$ physics anomalies as mentioned earlier. 
In order to get a large $BR(\Delta^{\pm2/3} \to t N_\tau)$, only large values of
$Y^{33}_R$ are allowed which nearly saturates the perturbative limits $\sim 4\pi$ even at the TeV scale. 
However, we recall that LQs are always associated with larger symmetry. Assuming a larger gauge group at the 
TeV scale can potentially help to avoid Landau-pole problem due to large value of $Y^{33}_R$ .    

\begin{table}[h!]
\begin{center}
\begin{tabular}{||c|c|c|c||} 
\hline 
Parameters & BP1 & BP2 & BP3  \\
\hline\hline  
${m_D}^{ii}$ &  (0.1, 0.1, 0.1) & $(10^{-9}, 10^{-9}, 0.1)$ & (0.1, 0.1, 0.1) \\
\hline
${M_R}^{ii}$ (GeV) &  (1000.0, 1000.0, 400.0) & $(10^{-6}, 10^{-6}, 600)$  & (1000.0, 1000.0, 800.0)\\ 
\hline
$\mu_X$ (eV) & {\tiny $\left(\begin{array}{ccc}
9.233 & 15.141 & 2.799\\
15.141 &  52.874& 22.228\\
2.799 & 22.228 & 15.921\\
\end{array}\right)$} & {\tiny $\left(\begin{array}{ccc}
0.144 & 0.237 & 0.525\\
0.237 & 0.826 & 4.168\\
0.525 & 4.168 & 35.821\\
\end{array}\right)$} & {\tiny $\left(\begin{array}{ccc}
14.427 & 23.658  & 6.998\\
23.658 & 82.616 & 55.569\\
6.998 & 55.569 & 63.683\\ 
\end{array}\right)$}\\
\hline\hline 
Resulting PMNS & {\tiny $\left(\begin{array}{ccc}
0.810 & 0.507 & 0.295\\
0.567 &  0.550 & 0.612\\
0.148 & 0.663 & 0.733\\
\end{array}\right)$} & {\tiny $\left(\begin{array}{ccc}
0.806 & 0.493 & 0.284\\
0.548 &  0.556 & 0.610\\
0.143 & 0.647 & 0.740\\
\end{array}\right)$}& {\tiny $\left(\begin{array}{ccc}
0.806 & 0.511  & 0.297\\
0.573 & 0.549 & 0.608\\
0.148 & 0.661 & 0.736\\
\end{array}\right)$}\\ 
\hline\hline 
\end{tabular}
\caption{$m_D$ and $M_R$ taken as inputs, $\mu_X$ is the resulting matrix derived such that the oscillation parameters 
are in agreement.}.
\label{tab:param} 
\end{center}
\end{table} 
Table~\ref{tab:param} shows the choices of the neutrino sector parameters, 
$m_D$ and $M_R$ and the resulting $\mu_X$ after fitting the neutrino oscillation data as mentioned in Section~\ref{subsec:numass}. 
We have also presented the obtained light $3\times 3$ neutrino mass matrices which are in good agreement with the allowed PMNS 
matrix elements. Non-diagonal structure of $\mu_X$ reflects to the fact that we have assumed diagonal $Y_\nu^{ii}$, hence $m_D^{ii}$ to suppress lepton flavor violating obsevables. In the present context, the effective production cross-section of the right handed neutrinos have been considerably larger, thanks to 
larger productions for LQs ($\Delta^{(2/3)}$) and their decays to $t N_\tau$ with almost 100\% branching ratio.
\begin{table}[h!]
\begin{center}
\begin{tabular}{||c|c|c|c|c|c||} 
\hline 
Benchmarks  & $m_{\Delta^{(2/3)}}$ (GeV) & $m_{N_\tau}$ (GeV) & ${Y_L}^{ii}$ & ${Y_R}^{ii}$ & $\sigma_{\Delta^{(2/3)} \bar{\Delta}^{(2/3)}}$ (fb)  \\ 
\hline\hline 
BP1 & 850.0 &  400.0  & $(0.0, 10^{-3}, 10^{-3}$ & $(0.0, 10^{-3}, 0.1)$ & 18.760  \\
BP2 & 1000.0 & 600.0  & ($0.0$, $10^{-3}$, 1.5) & ($0.0$, $10^{-3}$, 12.56) & 6.342 \\
BP3 & 1200.0 & 800.0 & $(0.0, 10^{-3}, 10^{-3}$ & $(0.0, 10^{-3}, 0.1)$ & 1.512  \\
\hline\hline
\end{tabular}
\caption{Relevant parameters and production cross-section (including the k-factor) of leptoquarks at LHC, for 13 TeV c.m.energy.}
\label{tab:bp}
\end{center}
\end{table} 
Table~\ref{tab:bp} shows our choices of LQ ($\Delta^{(2/3)}$) masses and their relevant couplings along with their 
pair production cross-section for the three benchmark points. 
We now proceed to discuss our collider analysis of the leptoquarks decaying via heavy neutrinos yielding various possible 
novel signal regions. 
\section{Collider Analysis and results
\label{sec:collans} }
As has already been discussed, our primary focus is on the interaction induced by the 
Yukawa couplings of right-handed neutrinos with the leptoquarks which give rise to  interesting signals at the LHC. 
Leptoquarks, being strongly interacting, are copiously produced at hadron colliders. Therefore, in the framework of 
leptoquark model, the production of right handed neutrinos could be enhanced significantly. Moreover, 
the decay of leptoquarks are usually accompanied by hard jets and/or leptons which could enhance 
the efficiency of signal selection criteria to reduce the SM background. With the LHC running at its near kinematic threshold, 
at $\sqrt{s}=13$ TeV, one also has to deal with the large hadronic backgrounds while looking for  any new physics signal. 
One way to avoid this menace is to look for more and more lepton enriched final states which naturally tend to reduce these 
unwanted background contributions. 
 We have, therefore, focussed to identify the signal regions in a way so that the maximum
number of hard leptons, like-sign or otherwise, are identified in the final state while not tagging all the $b$-jets and $\tau$-jets arising from the 
cascades in order to maximize the signal rates.                              
In the framework of the present model, different signal regions (SR) can be defined with multiple leptons when 
$\Delta^{\pm2/3}\to t N_\tau$ and $N_\tau \to \tau W ~{\rm or}~ \nu_{\tau} Z$ are followed by the leptonic decay of both the top quarks and the
gauge bosons present in the cascade. 
\begin{equation}
\begin{split}
pp & \to \Delta^{\pm2/3}\Delta^{\mp2/3}  \to  (t \bar N_\tau)(\bar t N_\tau) \\ 
&\to b\bar b W^+ W^-\tau^+\tau^-W^+W^- \\
&\to b\bar b W^+ W^-\nu_{\tau}\bar\nu_{\tau}ZZ \\
&\to b\bar b W^+ W^-\tau^+\nu_{\tau}W^-Z
\end{split}
\label{eq:4l}
\end{equation}
The above three possible decay chains can potentially give rise to maximum six leptons in the final state. Any final state 
with such a large lepton multiplicity will be a very clean signal devoid of SM background. However, 
that will require both the $N_\tau$s to decay via the Z-boson which has a suppressed branching ratio compared to 
its W-decay mode. Moreover, the $Z$-boson decay branching ratios to the electrons and muons are also suppressed. 
Hence we restrict ourselves to signal regions with at least three or four leptons in the final state. Note that, 
here we focus on the RH neutrino decay to $\tau$ lepton. The $e, \mu$ states are mostly generated from the gauge 
boson decays. Another interesting signal region (SR) can be constructed from the first and third cascade decays 
shown in Eq.~\ref{eq:4l} when two of the same-sign $W$-bosons decay leptonically and the other $W$ and (or)
$Z$-bosons decay hadronically. This will result into a same-sign dilepton (SSD) signal along with multiple jets and $\met$. 
Such a SSD signal is expected to be different from the usual heavy neutrino signals both in terms of jet multiplicity 
and $\met$ distribution. 

For a detailed collider simulation, we have generated parton level events using MadGraph5 and subsequently 
passed the events into Pythia (v6.4.28) \cite{Sjostrand:2006za} for simulating initial state radiation/final state 
radiation (ISR/FSR), decay of the heavy particles and hadronisation of the final state quarks. In order to reconstruct 
the physics objects like, jets, leptons, photons and missing transverse energy, we have used the fast detector simulator 
Delphes-v3.3.3 \cite{deFavereau:2013fsa,Selvaggi:2014mya,Mertens:2015kba}. The jets are reconstructed by 
anti-$k_t$ algorithm \cite{Cacciari:2008gp} implemented in Fastjet package \cite{Cacciari:2005hq,Cacciari:2008gp,Cacciari:2011ma} 
with a cone of $\Delta R=0.4$ and minimum transverse momentum of 20 GeV. 
The tagging and mistagging efficiencies of $b$-jet and $\tau$-jet have been incorporated according to the latest 
ATLAS studies in this regard \cite{ATLAS:2015-022} in Delphes3.
After the object reconstruction, leptons with $p_T^l>10$ GeV and $|\eta^l|<2.5$ and jets with 
$p_T^j>20$ GeV and $|\eta^j|<2.5$ are considered for the further event selection.  
Furthermore, the electrons and muons satisfying the $p_T^l$ criteria have been selected with 95\% and 85\% efficiencies 
for $|\eta^l|<1.5$ and $1.5<|\eta^l|<2.5$ respectively.
 Finally, the $b$ and $\tau$ jets are tagged with $p_T>20$ GeV while the rest of the jets are tagged with 
different $p_T$ requirements of our signal regions following the ATLAS event selection criteria as in 
Refs.~\cite{Aaboud:2017dmy,Aad:2016tuk}. 

\begin{figure}[h!]
\begin{center}
\includegraphics[width=7.0cm,height=7.0cm]{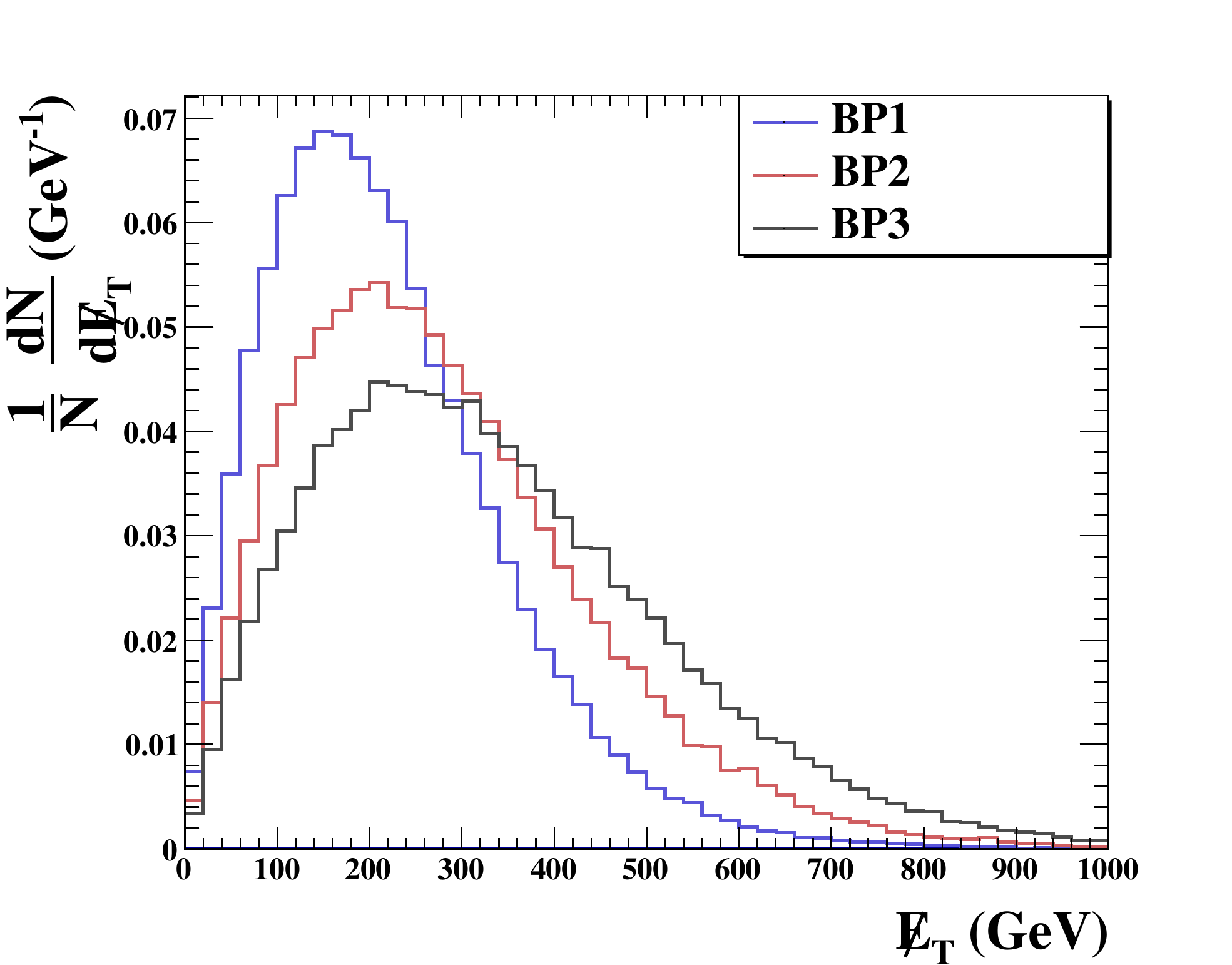}
\includegraphics[width=7.0cm,height=7.0cm]{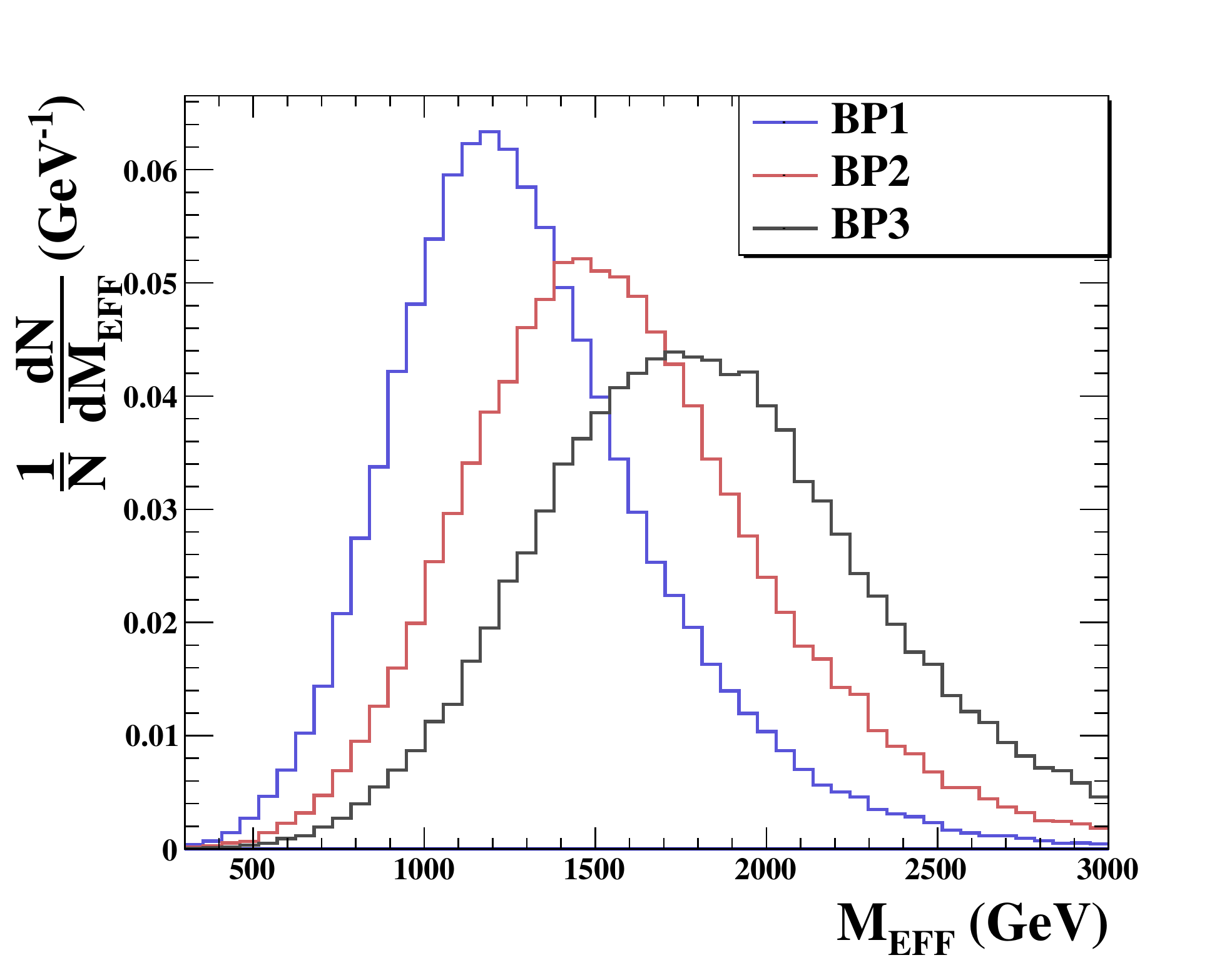}
\caption{Normalised distributions of $\met$ and $M_{EFF}$ for final states consisting of at least two leptons corresponding to 
the three benchmark points.}
\label{fig:met_meff}
\end{center}
\end{figure}

Multiple sources of missing energy throughout the cascade and the large multiplicity of both jets and(or) leptons 
are likely to produce hard $\met$ and $M_{EFF}$ distributions, where $M_{EFF} = \sum_i p^j_{T_i} + \sum_i p^{\ell}_{T_i} + \met$.
In Fig.~\ref{fig:met_meff}, we have shown these kinematic distributions for the three benchmark points 
introduced in Section~\ref{sec:stsconstbps}.

The distributions in Fig.~\ref{fig:met_meff} are obtained for final states containing at least two leptons. 
The blue, red and black lines represent BP1, BP2 and BP3 respectively. As expected, the hardness of the distributions 
increases with increasing leptoquark and heavy neutrino masses. The final state should also consist a large number of 
jets including multiple $b$ and $\tau$-jets. Tagging the $b/\tau$-jets can reduce the SM backgrounds significantly. 
The large QCD background at the LHC makes it difficult for any new physics signal with multiple jets to be observed. Tagging 
additional $b/\tau$-jets helps to reduce this background very effectively. Top-quark associated production channels also give rise to 
large background contributions which can be minimised by demanding a $\tau$-jet in the final states.
However, the detection efficiencies of $b/\tau$-jets vary with their $p_T$s and hence tagging all the $b$ and (or) $\tau$-jets can also reduce 
the signal rates considerably. Taking into account all these factors, we define four SRs as shown in Table~\ref{tab:SRs}. 
\begin{table}[h!]
\begin{center}
\begin{tabular}{||c||c|c|c|c|c|c|c|c||} 
\hline
\hline
\multicolumn{1}{||c||}{Signal} &
\multicolumn{6}{|c|}{\bf C0} & 
\multicolumn{1}{|c|}{\bf C1} & 
\multicolumn{1}{|c|}{\bf C2} \\
\cline{2-9}
Region & $N_{\ell}$ & $N_{j}$ & $N_{b}$ & $N_{\tau}$ & $p_T^{j}$ (GeV) & $p_T^{b/\tau}$ (GeV) & $\met$ (GeV) & $M_{EFF}$ (GeV)  \\
\hline\hline 
{\bf SR3Lb}   & $\ge 3$ & $\ge 4$  & $\ge 1$ & - & $> 40$ & $> 20$ & $> 200$ & $> 600$  \\
{\bf SR3Lbt}  & $\ge 3$ & $\ge 4$  & $\ge 1$ & $\ge 1$ & $> 40$ & $> 20$ & $> 200$ & $> 600$  \\
{\bf SR4Lbt}  & $\ge 4$ & -  & $\ge 1$ & $\ge 1$ & $> 40$ & $> 20$ & $> 200$ & $> 600$  \\
{\bf SRSS2Lb}  & $\ge 2$ & $\ge 4$  & $\ge 1$ & -  & $> 50$ & $> 20$ & $> 150$ & $\ge 550$  \\
\hline\hline
\end{tabular}
\caption{The four different signal regions with lepton, jet multiplicities and kinematic cuts.}
\label{tab:SRs}
\end{center}
\end{table}
We have categorised the choice of lepton and jet multiplicities and their $p_T$ requirement of the jets, $\met$ and $M_{EFF}$ under 
{\bf C0}, {\bf C1} and {\bf C2} respectively as shown in the table.

 For most of the signal regions focus on the large 
lepton and jet multiplicity and the hard $\met$ and $M_{EFF}$ distributions to achieve a good signal to background ratio.
Among the first three SRs, {\bf SR3Lb} and {\bf SR3Lbt} are expected to have some SM background contributions while 
{\bf SR4Lbt} is likely to be a much cleaner channel. As mentioned before, any lepton number violating signals are 
suppressed in the framework of the inverse seesaw scenario 
due to the smallness of $\mu_X$. Hence the usual SSD signal is not viable to probe for inverse seesaw extended SM.   
However, here one can obtain a different kind of same-sign dilepton signal as shown in {\bf SRSS2Lb}, where
the same-sign criteria is imposed on the events with exactly two leptons in the final state. Such a signal region can, therefore, be 
a distinguishing feature for such a scenario.
The choice of our signal regions are motivated from some experimental studies with final states consisting of 
multiple jets and leptons in the context of Supersymmetry \cite{Aad:2014pda,Aaboud:2017dmy,Aad:2016tuk}. There are multiple 
signal regions studied in these works, which can be relevant to our scenario. Therefore, we have checked the consistency of 
our chosen benchmark points with these experimental results using CheckMATE \cite{Drees:2013wra,Dercks:2016npn} which determines 
whether a chosen model parameter space is excluded or not at 95$\%$ confidence level \cite{Read:2002hq} by comparing it to the 
relevant experimental analyses.\footnote{The analyses of Ref.\cite{Aaboud:2017dmy} have not yet been included in CheckMATE. We 
have checked consistency of our benchmark points with these results using our own code.}

The obtained signal cross-sections at 13 TeV center-of-mass energy 
are presented in Table~\ref{tab:bp_sig} with cut-flow numbers for the signal regions and cuts {\bf C0} - {\bf C2} as defined 
in Table~\ref{tab:SRs} corresponding to the three BPs defined in Table~\ref{tab:bp}. Irrespective of the choice of different signal 
regions, the $M_{EFF}$ cut ({\bf C2}) does not affect the signal cross-sections for any of the benchmark points. This is quite expected 
as Fig.~\ref{fig:met_meff} clearly shows our choice of the $\met$ cut ({\bf C1}) automatically pushes the phase space to a hard kinematic 
region where the $M_{EFF}$ cut is trivially satisfied. However, {\bf C2} is very useful to reduce the SM backgrounds for these signal 
regions as mentioned in \cite{Aaboud:2017dmy,Aad:2016tuk}.
\begin{table}[h!]
\begin{center}
\begin{tabular}{||c||c|c|c||c|c|c||c|c|c||} 
\hline
\multicolumn{1}{||c||}{} &
\multicolumn{9}{c||}{Signal cross-section (fb)} \\
\cline{2-10}
\multicolumn{1}{||c||}{Signal} &
\multicolumn{3}{c||}{\bf BP1} &
\multicolumn{3}{c||}{\bf BP2} &
\multicolumn{3}{c||}{\bf BP3} \\
\cline{2-10}
Region & {\bf C0} & {\bf C1} & {\bf C2} & {\bf C0} & {\bf C1} & {\bf C2} & {\bf C0} & {\bf C1} & {\bf C2}  \\
\hline\hline 
{\bf SR3Lb}   & 0.376 & 0.181 & 0.181 & 0.130 & 0.081 & 0.081 & 0.330 & 0.024 & 0.024 \\
\hline
{\bf SR3Lbt}  & 0.137 & 0.062 & 0.062 & 0.047 & 0.027 & 0.027 & 0.012 & 0.009 & 0.009 \\
\hline
{\bf SR4Lbt}  & 0.016 & 0.007 & 0.007 & 0.006 & 0.003 & 0.003 & 0.002 & 0.001 & 0.001 \\
\hline
{\bf SRSS2Lb}  & 0.924 & 0.598 & 0.598 & 0.320 & 0.245 & 0.245 & 0.080 & 0.066 & 0.066  \\
\hline\hline
\end{tabular}
\caption{Cut-flow table for signal cross-sections with the cuts mentioned in Table~\ref{tab:SRs} for the different SRs and BPs at the LHC at 13 TeV center-of-mass energy.}
\label{tab:bp_sig}
\end{center}
\end{table}

\begin{itemize}
\item
The signal region {\bf SR3Lb} with three or more leptons and at least one tagged $b$-jet has also been studied 
by the ATLAS Collaboration although in a different context \cite{Aaboud:2017dmy}. They estimated a combined SM 
background cross-section of 0.56 fb. In absence of any significant deviation from the SM background prediction, 
the ATLAS collaboration has derived an upper bound of 0.41 fb on new physics tri-lepton cross-section in {\bf SR3Lb} 
signal region. As the numbers in Table~\ref{tab:bp_sig} suggest, all our benchmark points are well within this 
experimental bound. We have also checked that the ATLAS study of {\bf SR3Lb} imposes a lower bound of about 700 GeV 
on the mass of the 3rd generation leptoquark if it dominantly decays into a top-quark and a heavy neutrino in the 
framework of the present model. Taking into account the SM background contribution, one can 
obtain a statistical significance\footnote{Statistical significance, $\mathcal{S}=\frac{s}{\sqrt{s+b}}$, where 
$s$=number of signal events and $b$=number of background events.} of $3\sigma$ for BP1 at an integrated luminosity ($L$)
of $\sim 200~{\rm fb}^{-1}$. For BP2, with higher LQ mass, in order to achieve similar statistical significance, 
one needs $L\sim 900~{\rm fb}^{-1}$. BP3 is unlikely to be probed in this SR even at a relatively high luminosity of 
$L\sim 3000~{\rm fb}^{-1}$.

\item
Demanding at least one additional $\tau$-jet {({\em i.e.,} the SR {\bf SR3Lbt})} reduces the signal cross-section to 
0.062 fb  for BP1.  However, demanding an extra $\tau$-tagged jet on the top of {\bf SR3Lb} reduces the SM background 
cross-section at least by a factor of $\alpha_W$. Thus, even the $\sim 12$ events 
expected at $L\sim 200~{\rm fb}^{-1}$ for BP1 is relatively background free and can act as a complementary channel to {\bf SR3Lb}.
This SR can be more economical for BP2 and BP3 compared to the previous one. As the numbers in Table~\ref{tab:bp_sig} suggest, 
both these benchmark points can be probed below and around $L=1000~{\rm fb}^{-1}$ respectively. 

\item
{\bf SR4Lbt} is even a more cleaner channel to probe for the present scenario. However, as the numbers suggest, in order to get 
some signal events ($N_{sig}\sim 7$ for BP1) in this signal region, one has to go for high luminosity $L\sim 1000~{\rm fb}^{-1}$. 
Thus this signal region can only serve as a complementary channel to the previous two if any excess of events are found in either of them. 
For LQ masses around 1000 GeV and above, as in BP2 and BP3, this SR does not look promising enough due to extremely small signal 
cross-section.

\item
 The signal region {\bf SRSS2Lb} has been studied in the context of supersymmetry at the LHC. Here we have used 
the same set of selection criteria and kinematic cuts as the ATLAS collaboration \cite{Aad:2016tuk} to obtain the 
signal cross-section. The experimental upper limit on new physics cross-section in this signal region remains at 
2.8 fb which is much larger than that obtained for any of the benchmark points of our choice.  
Although the SM background contribution in this channel is expected to be the largest compared to the other {\bf SR}s, the large 
signal cross-section makes it the most promising channel to probe at the LHC. The ATLAS collaboration quoted the total SM 
background contribution as 1.406 fb \cite{Aad:2016tuk}. This leads to a statistical significance of $3\sigma$ at 
$L\sim 55~{\rm fb}^{-1}$ for BP1 which means this parameter space can immediately be probed with the accumulated data at 
the LHC at present. To achieve similar statistical significance for BP2 and BP3 one requires $L\sim 250~{\rm fb}^{-1}$ and 
$3000~{\rm fb}^{-1}$ respectively. 

\end{itemize} 

From the above, it is evident that a lighter LQ mass and lighter RH neutrino, such as $M_{\Delta^{2/3}}=850, m_{N_\tau}=400$ GeV  
can be probed in very near future with the data already accumulated at the LHC 
through at least two same-sign leptons, multi-jet and one tagged $b$-jet in the final state. Even a heavier neutrino ($\sim 600$ GeV) can be probed in association with a 
$\sim 1000$ GeV leptoquark mass at a relatively lower luminosity of $250~{\rm fb}^{-1}$ in this final state. To probe higher mass 
ranges, as represented by BP3, the HL-LHC run is required. Among the other SRs, {\bf SR3Lb} and {\bf SR3Lbt} turn out to 
be the other viable options. However, they will be relevant only if some hint of new physics is obtained at lower luminosity 
via {\bf SRSS2Lb}. Besides, it also provides us with a nice distinguishing feature that can be used to differentiate an usual inverse 
seesaw extended SM scenario from the present one. Note that, almost two third contribution of the signal cross-section in this SR 
arises from same-sign two-lepton final state, a contribution which is expected to be negligible in the usual inverse seesaw extended SM. 

\section{Conclusion \& Outlook}
\label{sec:concl}
In this work, we have considered a minimal extension of the SM that can (i) produce neutrino masses and mixing angles and 
(ii) explain $B-$ physics anomalies. We consider a scalar LQ  with the hypercharge 
$Y=\frac{1}{6}$ and embed it in a inverse-seesaw framework with TeV scale or even lighter sterile neutrinos. Presence of a 
dominant flavor diagonal LQ-$\nu_R$ coupling can greatly enhance the effective production cross-section of the heavy neutrino states in comparison to the canonical sterile neutrino production through charged/neutral current weak interactions. 
  We have studied a rather unexplored decay mode of the leptoquark, i.e, to a top quark and heavy right handed neutrino  with almost 100\% Br. This results 
 in various multi lepton signals associated with multiple jets (including $b$ and $\tau$-jets) and missing transverse energy.
We have explored four such signal regions and carried out a comparative analysis aimed at probing heavy neutrino masses. 
We find that  while a RH neutrino of mass 400 GeV can immediately be probed at the LHC with at least one same-sign 
dilepton and multi-jet signal, a relatively heavier 
neutrino mass scales such as 800 GeV can be  probed in the high luminosity run of LHC. We have also observed 
that a mass bound of 700 GeV on the LQ mass can be derived following the ATLAS 13 TeV search in trilepton channel 
associated with multiple jets and at least one b-tagged jet provided the LQ decays dominantly into a top-quark and a heavy 
neutrino.   

We have also presented a same-sign dilepton signal region which is expected to yield a much smaller 
event rate at the collider in the inverse seesaw extended SM. However, in the presence of leptoquarks where the heavy neutrinos 
are produced from the decay of these colored particles alongside top quarks, we have shown that one can expect significant 
event rates in this same-sign dilepton channel associated with multiple jets and missing energy. 

Qualitatively all these can be considered as nice distinguishing features between the conventional inverse seesaw and the leptoquark 
associated inverse seesaw model. Quantatively, one can observe that much larger mass scale for heavy neutrinos can be reached through the aforesaid signal regions, specially via same-sign dileptons. Finally, allowing small off-diagonal structure in the LQ-$\nu_R$ coupling may lead to 
new avenues to LHC physics and flavor violations which requires a dedicated study. 
\section*{Acknowledgments}
KG and MM acknowledges the support of DST-INSPIRE FACULTY research grant. 
The work of SM is partially supported by funding available from the Department of Atomic Energy, Government of India, 
for the Regional Centre for Accelerator-based Particle Physics (RECAPP), Harish-Chandra Research Institute. SM 
acknowledges the hospitality of Institute of Physics, Bhubaneswar during initial phase of the work. 
\section*{Appendix}
\subsection*{Relevant decays of the Leptoquark: }
Below, we write down the different two body decays of the LQ, relevant for our work.
\begin{align}
&\Gamma(\Delta^{2/3}\to t\bar N_{\tau}) = \frac{(Y_R^{33})^2 ( (V_N^{53})^2 + (V_N^{63})^2 )}{16\pi(m_{\Delta^{(2/3)}})^3} ((m_{\Delta^{(2/3)}})^2 
- m_{N_{\tau}}^2 - m_t^2)\times \cr  
&\sqrt{((m_{\Delta^{(2/3)}})^2 - (m_{N_{\tau}} + m_t)^2)((m_{\Delta^{(2/3)}})^2 - (m_{N_{\tau}} - m_t)^2)} \\
&\Gamma(\Delta^{2/3}\to b\bar\tau) = \frac{(Y_L^{33})^2}{16\pi(m_{\Delta^{(2/3)}})^3} ( (m_{\Delta^{(2/3)}})^2 - m_b^2 
- m_{\tau}^2)\times \cr  
&\sqrt{((m_{\Delta^{(2/3)}})^2 - (m_b + m_{\tau})^2)((m_{\Delta^{(2/3)}})^2 - (m_b - m_{\tau})^2)} 
\end{align}
where, $V_N^{ij}$ ($i,j = 1,.....,6$) diagonalises the heavy neutrino mass matrix (${\mathcal M}_N$) written in the basis 
$\{\nu_R^1,\nu_R^2,\nu_R^3,X^1,X^2,X^3\}$.
Note that all the six heavy neutrino masses are driven by the choices of $M_R^{ii}$ since all the diagonal entries of ${\mathcal M}_N$ 
are essentially zero ($\mu_X << M_R$). Upon diagonalisation of ${\mathcal M}_N$, we obtain three pairs of mass degenerate heavy neutrino states 
with masses $\simeq M_R^{11}$, $M_R^{22}$ and $M_R^{33}$ respectively.  
While writing $V_N$, the eigenvalues are arranged from heavy to light order. The fifth and sixth eigenvalues (mass degenerate) are driven by the  
chice of $M_R^{33}$ since by our choice, $M_R^{33} < M_R^{11}, M_R^{22}$. Only $V_N^{53}$ and $V_N^{63}$ elements are of importance for the 
decay $\Delta^{2/3}\to t\bar N_{\tau}$ since the relevant 
term in the Lagrangian is $Y_R^{ij} \bar u_L^{i} \nu_R^{j} \Delta^{(2/3)}$ and we have chosen to work with diagonal $Y_R$. 
$m_t$, $m_b$ and $m_{\tau}$ are the masses of top quark, bottom quark and $\tau$-lepton respectively. 
\bibliography{lq_invs}{}

\begin{thebibliography}{101}%
\makeatletter
\providecommand \@ifxundefined [1]{%
 \@ifx{#1\undefined}
}%
\providecommand \@ifnum [1]{%
 \ifnum #1\expandafter \@firstoftwo
 \else \expandafter \@secondoftwo
 \fi
}%
\providecommand \@ifx [1]{%
 \ifx #1\expandafter \@firstoftwo
 \else \expandafter \@secondoftwo
 \fi
}%
\providecommand \natexlab [1]{#1}%
\providecommand \enquote  [1]{``#1''}%
\providecommand \bibnamefont  [1]{#1}%
\providecommand \bibfnamefont [1]{#1}%
\providecommand \citenamefont [1]{#1}%
\providecommand \href@noop [0]{\@secondoftwo}%
\providecommand \href [0]{\begingroup \@sanitize@url \@href}%
\providecommand \@href[1]{\@@startlink{#1}\@@href}%
\providecommand \@@href[1]{\endgroup#1\@@endlink}%
\providecommand \@sanitize@url [0]{\catcode `\\12\catcode `\$12\catcode
  `\&12\catcode `\#12\catcode `\^12\catcode `\_12\catcode `\%12\relax}%
\providecommand \@@startlink[1]{}%
\providecommand \@@endlink[0]{}%
\providecommand \url  [0]{\begingroup\@sanitize@url \@url }%
\providecommand \@url [1]{\endgroup\@href {#1}{\urlprefix }}%
\providecommand \urlprefix  [0]{URL }%
\providecommand \Eprint [0]{\href }%
\providecommand \doibase [0]{http://dx.doi.org/}%
\providecommand \selectlanguage [0]{\@gobble}%
\providecommand \bibinfo  [0]{\@secondoftwo}%
\providecommand \bibfield  [0]{\@secondoftwo}%
\providecommand \translation [1]{[#1]}%
\providecommand \BibitemOpen [0]{}%
\providecommand \bibitemStop [0]{}%
\providecommand \bibitemNoStop [0]{.\EOS\space}%
\providecommand \EOS [0]{\spacefactor3000\relax}%
\providecommand \BibitemShut  [1]{\csname bibitem#1\endcsname}%
\let\auto@bib@innerbib\@empty
\bibitem [{\citenamefont
  {{\url{https://twiki.cern.ch/twiki/bin/view/AtlasPublic}}}()}]{lhc1}%
  \BibitemOpen
  \bibfield  {author} {\bibinfo {author} {\bibnamefont
  {{\url{https://twiki.cern.ch/twiki/bin/view/AtlasPublic}}}},\ }\href@noop {}
  {\ }\BibitemShut {NoStop}%
\bibitem [{\citenamefont
  {{\url{https://twiki.cern.ch/twiki/bin/view/CMSPublic}}}()}]{lhc2}%
  \BibitemOpen
  \bibfield  {author} {\bibinfo {author} {\bibnamefont
  {{\url{https://twiki.cern.ch/twiki/bin/view/CMSPublic}}}},\ }\href@noop {} {\
  }\BibitemShut {NoStop}%
\bibitem [{\citenamefont {Gonzalez-Garcia}\ and\ \citenamefont
  {Maltoni}(2008)}]{GonzalezGarcia:2007ib}%
  \BibitemOpen
  \bibfield  {author} {\bibinfo {author} {\bibfnamefont {M.~C.}\ \bibnamefont
  {Gonzalez-Garcia}}\ and\ \bibinfo {author} {\bibfnamefont {M.}~\bibnamefont
  {Maltoni}},\ }\href {\doibase 10.1016/j.physrep.2007.12.004} {\bibfield
  {journal} {\bibinfo  {journal} {Phys. Rept.}\ }\textbf {\bibinfo {volume}
  {460}},\ \bibinfo {pages} {1} (\bibinfo {year} {2008})},\ \Eprint
  {http://arxiv.org/abs/0704.1800} {arXiv:0704.1800 [hep-ph]} \BibitemShut
  {NoStop}%
\bibitem [{\citenamefont {Gonzalez-Garcia}\ \emph {et~al.}(2016)\citenamefont
  {Gonzalez-Garcia}, \citenamefont {Maltoni},\ and\ \citenamefont
  {Schwetz}}]{Gonzalez-Garcia:2015qrr}%
  \BibitemOpen
  \bibfield  {author} {\bibinfo {author} {\bibfnamefont {M.~C.}\ \bibnamefont
  {Gonzalez-Garcia}}, \bibinfo {author} {\bibfnamefont {M.}~\bibnamefont
  {Maltoni}}, \ and\ \bibinfo {author} {\bibfnamefont {T.}~\bibnamefont
  {Schwetz}},\ }\href {\doibase 10.1016/j.nuclphysb.2016.02.033} {\bibfield
  {journal} {\bibinfo  {journal} {Nucl. Phys.}\ }\textbf {\bibinfo {volume}
  {B908}},\ \bibinfo {pages} {199} (\bibinfo {year} {2016})},\ \Eprint
  {http://arxiv.org/abs/1512.06856} {arXiv:1512.06856 [hep-ph]} \BibitemShut
  {NoStop}%
\bibitem [{\citenamefont {Minkowski}(1977)}]{Minkowski:1977sc}%
  \BibitemOpen
  \bibfield  {author} {\bibinfo {author} {\bibfnamefont {P.}~\bibnamefont
  {Minkowski}},\ }\href {\doibase 10.1016/0370-2693(77)90435-X} {\bibfield
  {journal} {\bibinfo  {journal} {Phys. Lett.}\ }\textbf {\bibinfo {volume}
  {B67}},\ \bibinfo {pages} {421} (\bibinfo {year} {1977})}\BibitemShut
  {NoStop}%
\bibitem [{\citenamefont {Mohapatra}\ and\ \citenamefont
  {Senjanovic}(1980)}]{Mohapatra:1979ia}%
  \BibitemOpen
  \bibfield  {author} {\bibinfo {author} {\bibfnamefont {R.~N.}\ \bibnamefont
  {Mohapatra}}\ and\ \bibinfo {author} {\bibfnamefont {G.}~\bibnamefont
  {Senjanovic}},\ }\href {\doibase 10.1103/PhysRevLett.44.912} {\bibfield
  {journal} {\bibinfo  {journal} {Phys. Rev. Lett.}\ }\textbf {\bibinfo
  {volume} {44}},\ \bibinfo {pages} {912} (\bibinfo {year} {1980})}\BibitemShut
  {NoStop}%
\bibitem [{\citenamefont {Gell-Mann}\ \emph {et~al.}()\citenamefont
  {Gell-Mann}, \citenamefont {Ramond},\ and\ \citenamefont
  {Slansky}}]{GellMann:1980vs}%
  \BibitemOpen
  \bibfield  {author} {\bibinfo {author} {\bibfnamefont {M.}~\bibnamefont
  {Gell-Mann}}, \bibinfo {author} {\bibfnamefont {P.}~\bibnamefont {Ramond}}, \
  and\ \bibinfo {author} {\bibfnamefont {R.}~\bibnamefont {Slansky}},\
  }\href@noop {} {\ }\bibinfo {note} {Print-80-0576 (CERN)}\BibitemShut
  {NoStop}%
\bibitem [{\citenamefont {Yanagida}()}]{Yanagida:1979as}%
  \BibitemOpen
  \bibfield  {author} {\bibinfo {author} {\bibfnamefont {T.}~\bibnamefont
  {Yanagida}},\ }\href@noop {} {\ }\bibinfo {note} {In Proceedings of the
  Workshop on the Baryon Number of the Universe and Unified Theories, Tsukuba,
  Japan, 13-14 Feb 1979}\BibitemShut {NoStop}%
\bibitem [{\citenamefont {Glashow}(1980)}]{Glashow:1979nm}%
  \BibitemOpen
  \bibfield  {author} {\bibinfo {author} {\bibfnamefont {S.~L.}\ \bibnamefont
  {Glashow}},\ }\href@noop {} {\bibfield  {journal} {\bibinfo  {journal} {NATO
  Adv. Study Inst. Ser. B Phys.}\ }\textbf {\bibinfo {volume} {59}},\ \bibinfo
  {pages} {687} (\bibinfo {year} {1980})}\BibitemShut {NoStop}%
\bibitem [{\citenamefont {Schechter}\ and\ \citenamefont
  {Valle}(1982)}]{Schechter:1981cv}%
  \BibitemOpen
  \bibfield  {author} {\bibinfo {author} {\bibfnamefont {J.}~\bibnamefont
  {Schechter}}\ and\ \bibinfo {author} {\bibfnamefont {J.~W.~F.}\ \bibnamefont
  {Valle}},\ }\href {\doibase 10.1103/PhysRevD.25.774} {\bibfield  {journal}
  {\bibinfo  {journal} {Phys. Rev.}\ }\textbf {\bibinfo {volume} {D25}},\
  \bibinfo {pages} {774} (\bibinfo {year} {1982})}\BibitemShut {NoStop}%
\bibitem [{\citenamefont {Schechter}\ and\ \citenamefont
  {Valle}(1980)}]{Schechter:1980gr}%
  \BibitemOpen
  \bibfield  {author} {\bibinfo {author} {\bibfnamefont {J.}~\bibnamefont
  {Schechter}}\ and\ \bibinfo {author} {\bibfnamefont {J.~W.~F.}\ \bibnamefont
  {Valle}},\ }\href {\doibase 10.1103/PhysRevD.22.2227} {\bibfield  {journal}
  {\bibinfo  {journal} {Phys. Rev.}\ }\textbf {\bibinfo {volume} {D22}},\
  \bibinfo {pages} {2227} (\bibinfo {year} {1980})}\BibitemShut {NoStop}%
\bibitem [{\citenamefont {Weinberg}(1979)}]{Weinberg:1979sa}%
  \BibitemOpen
  \bibfield  {author} {\bibinfo {author} {\bibfnamefont {S.}~\bibnamefont
  {Weinberg}},\ }\href {\doibase 10.1103/PhysRevLett.43.1566} {\bibfield
  {journal} {\bibinfo  {journal} {Phys. Rev. Lett.}\ }\textbf {\bibinfo
  {volume} {43}},\ \bibinfo {pages} {1566} (\bibinfo {year}
  {1979})}\BibitemShut {NoStop}%
\bibitem [{\citenamefont {Weinberg}(1980)}]{Weinberg:1980bf}%
  \BibitemOpen
  \bibfield  {author} {\bibinfo {author} {\bibfnamefont {S.}~\bibnamefont
  {Weinberg}},\ }\href {\doibase 10.1103/PhysRevD.22.1694} {\bibfield
  {journal} {\bibinfo  {journal} {Phys. Rev.}\ }\textbf {\bibinfo {volume}
  {D22}},\ \bibinfo {pages} {1694} (\bibinfo {year} {1980})}\BibitemShut
  {NoStop}%
\bibitem [{\citenamefont {Magg}\ and\ \citenamefont
  {Wetterich}(1980)}]{Magg:1980ut}%
  \BibitemOpen
  \bibfield  {author} {\bibinfo {author} {\bibfnamefont {M.}~\bibnamefont
  {Magg}}\ and\ \bibinfo {author} {\bibfnamefont {C.}~\bibnamefont
  {Wetterich}},\ }\href {\doibase 10.1016/0370-2693(80)90825-4} {\bibfield
  {journal} {\bibinfo  {journal} {Phys. Lett.}\ }\textbf {\bibinfo {volume}
  {94B}},\ \bibinfo {pages} {61} (\bibinfo {year} {1980})}\BibitemShut
  {NoStop}%
\bibitem [{\citenamefont {Cheng}\ and\ \citenamefont
  {Li}(1980)}]{Cheng:1980qt}%
  \BibitemOpen
  \bibfield  {author} {\bibinfo {author} {\bibfnamefont {T.~P.}\ \bibnamefont
  {Cheng}}\ and\ \bibinfo {author} {\bibfnamefont {L.-F.}\ \bibnamefont {Li}},\
  }\href {\doibase 10.1103/PhysRevD.22.2860} {\bibfield  {journal} {\bibinfo
  {journal} {Phys. Rev.}\ }\textbf {\bibinfo {volume} {D22}},\ \bibinfo {pages}
  {2860} (\bibinfo {year} {1980})}\BibitemShut {NoStop}%
\bibitem [{\citenamefont {Foot}\ \emph {et~al.}(1989)\citenamefont {Foot},
  \citenamefont {Lew}, \citenamefont {He},\ and\ \citenamefont
  {Joshi}}]{Foot:1988aq}%
  \BibitemOpen
  \bibfield  {author} {\bibinfo {author} {\bibfnamefont {R.}~\bibnamefont
  {Foot}}, \bibinfo {author} {\bibfnamefont {H.}~\bibnamefont {Lew}}, \bibinfo
  {author} {\bibfnamefont {X.~G.}\ \bibnamefont {He}}, \ and\ \bibinfo {author}
  {\bibfnamefont {G.~C.}\ \bibnamefont {Joshi}},\ }\href {\doibase
  10.1007/BF01415558} {\bibfield  {journal} {\bibinfo  {journal} {Z. Phys.}\
  }\textbf {\bibinfo {volume} {C44}},\ \bibinfo {pages} {441} (\bibinfo {year}
  {1989})}\BibitemShut {NoStop}%
\bibitem [{\citenamefont {Antusch}\ and\ \citenamefont
  {Fischer}(2015)}]{Antusch:2015mia}%
  \BibitemOpen
  \bibfield  {author} {\bibinfo {author} {\bibfnamefont {S.}~\bibnamefont
  {Antusch}}\ and\ \bibinfo {author} {\bibfnamefont {O.}~\bibnamefont
  {Fischer}},\ }\href {\doibase 10.1007/JHEP05(2015)053} {\bibfield  {journal}
  {\bibinfo  {journal} {JHEP}\ }\textbf {\bibinfo {volume} {05}},\ \bibinfo
  {pages} {053} (\bibinfo {year} {2015})},\ \Eprint
  {http://arxiv.org/abs/1502.05915} {arXiv:1502.05915 [hep-ph]} \BibitemShut
  {NoStop}%
\bibitem [{\citenamefont {Mohapatra}(1986)}]{Mohapatra:1986aw}%
  \BibitemOpen
  \bibfield  {author} {\bibinfo {author} {\bibfnamefont {R.~N.}\ \bibnamefont
  {Mohapatra}},\ }\href {\doibase 10.1103/PhysRevLett.56.561} {\bibfield
  {journal} {\bibinfo  {journal} {Phys. Rev. Lett.}\ }\textbf {\bibinfo
  {volume} {56}},\ \bibinfo {pages} {561} (\bibinfo {year} {1986})}\BibitemShut
  {NoStop}%
\bibitem [{\citenamefont {Nandi}\ and\ \citenamefont
  {Sarkar}(1986)}]{Nandi:1985uh}%
  \BibitemOpen
  \bibfield  {author} {\bibinfo {author} {\bibfnamefont {S.}~\bibnamefont
  {Nandi}}\ and\ \bibinfo {author} {\bibfnamefont {U.}~\bibnamefont {Sarkar}},\
  }\href {\doibase 10.1103/PhysRevLett.56.564} {\bibfield  {journal} {\bibinfo
  {journal} {Phys. Rev. Lett.}\ }\textbf {\bibinfo {volume} {56}},\ \bibinfo
  {pages} {564} (\bibinfo {year} {1986})}\BibitemShut {NoStop}%
\bibitem [{\citenamefont {Mohapatra}\ and\ \citenamefont
  {Valle}(1986)}]{Mohapatra:1986bd}%
  \BibitemOpen
  \bibfield  {author} {\bibinfo {author} {\bibfnamefont {R.~N.}\ \bibnamefont
  {Mohapatra}}\ and\ \bibinfo {author} {\bibfnamefont {J.~W.~F.}\ \bibnamefont
  {Valle}},\ }\bibfield  {booktitle} {\emph {\bibinfo {booktitle}
  {{Proceedings, 23RD International Conference on High Energy Physics, JULY
  16-23, 1986, Berkeley, CA}}},\ }\href {\doibase 10.1103/PhysRevD.34.1642}
  {\bibfield  {journal} {\bibinfo  {journal} {Phys. Rev.}\ }\textbf {\bibinfo
  {volume} {D34}},\ \bibinfo {pages} {1642} (\bibinfo {year}
  {1986})}\BibitemShut {NoStop}%
\bibitem [{\citenamefont {'t~Hooft}(1980)}]{tHooft:1979rat}%
  \BibitemOpen
  \bibfield  {author} {\bibinfo {author} {\bibfnamefont {G.}~\bibnamefont
  {'t~Hooft}},\ }\bibfield  {booktitle} {\emph {\bibinfo {booktitle} {{Recent
  Developments in Gauge Theories. Proceedings, Nato Advanced Study Institute,
  Cargese, France, August 26 - September 8, 1979}}},\ }\href {\doibase
  10.1007/978-1-4684-7571-5_9} {\bibfield  {journal} {\bibinfo  {journal} {NATO
  Sci. Ser. B}\ }\textbf {\bibinfo {volume} {59}},\ \bibinfo {pages} {135}
  (\bibinfo {year} {1980})}\BibitemShut {NoStop}%
\bibitem [{\citenamefont {Cao}\ \emph {et~al.}(2017)\citenamefont {Cao},
  \citenamefont {Guo}, \citenamefont {He}, \citenamefont {Shang},\ and\
  \citenamefont {Yue}}]{Cao:2017cjf}%
  \BibitemOpen
  \bibfield  {author} {\bibinfo {author} {\bibfnamefont {J.}~\bibnamefont
  {Cao}}, \bibinfo {author} {\bibfnamefont {X.}~\bibnamefont {Guo}}, \bibinfo
  {author} {\bibfnamefont {Y.}~\bibnamefont {He}}, \bibinfo {author}
  {\bibfnamefont {L.}~\bibnamefont {Shang}}, \ and\ \bibinfo {author}
  {\bibfnamefont {Y.}~\bibnamefont {Yue}},\ }\href@noop {} {\  (\bibinfo {year}
  {2017})},\ \Eprint {http://arxiv.org/abs/1707.09626} {arXiv:1707.09626
  [hep-ph]} \BibitemShut {NoStop}%
\bibitem [{\citenamefont {Karmakar}\ and\ \citenamefont
  {Sil}(2017)}]{Karmakar:2016cvb}%
  \BibitemOpen
  \bibfield  {author} {\bibinfo {author} {\bibfnamefont {B.}~\bibnamefont
  {Karmakar}}\ and\ \bibinfo {author} {\bibfnamefont {A.}~\bibnamefont {Sil}},\
  }\href {\doibase 10.1103/PhysRevD.96.015007} {\bibfield  {journal} {\bibinfo
  {journal} {Phys. Rev.}\ }\textbf {\bibinfo {volume} {D96}},\ \bibinfo {pages}
  {015007} (\bibinfo {year} {2017})},\ \Eprint
  {http://arxiv.org/abs/1610.01909} {arXiv:1610.01909 [hep-ph]} \BibitemShut
  {NoStop}%
\bibitem [{\citenamefont {Sinha}\ \emph {et~al.}(2016)\citenamefont {Sinha},
  \citenamefont {Samanta},\ and\ \citenamefont {Ghosal}}]{Sinha:2015ooa}%
  \BibitemOpen
  \bibfield  {author} {\bibinfo {author} {\bibfnamefont {R.}~\bibnamefont
  {Sinha}}, \bibinfo {author} {\bibfnamefont {R.}~\bibnamefont {Samanta}}, \
  and\ \bibinfo {author} {\bibfnamefont {A.}~\bibnamefont {Ghosal}},\ }\href
  {\doibase 10.1016/j.physletb.2016.05.080} {\bibfield  {journal} {\bibinfo
  {journal} {Phys. Lett.}\ }\textbf {\bibinfo {volume} {B759}},\ \bibinfo
  {pages} {206} (\bibinfo {year} {2016})},\ \Eprint
  {http://arxiv.org/abs/1508.05227} {arXiv:1508.05227 [hep-ph]} \BibitemShut
  {NoStop}%
\bibitem [{\citenamefont {Arganda}\ \emph
  {et~al.}(2016{\natexlab{a}})\citenamefont {Arganda}, \citenamefont {Herrero},
  \citenamefont {Marcano},\ and\ \citenamefont {Weiland}}]{Arganda:2015naa}%
  \BibitemOpen
  \bibfield  {author} {\bibinfo {author} {\bibfnamefont {E.}~\bibnamefont
  {Arganda}}, \bibinfo {author} {\bibfnamefont {M.~J.}\ \bibnamefont
  {Herrero}}, \bibinfo {author} {\bibfnamefont {X.}~\bibnamefont {Marcano}}, \
  and\ \bibinfo {author} {\bibfnamefont {C.}~\bibnamefont {Weiland}},\ }\href
  {\doibase 10.1103/PhysRevD.93.055010} {\bibfield  {journal} {\bibinfo
  {journal} {Phys. Rev.}\ }\textbf {\bibinfo {volume} {D93}},\ \bibinfo {pages}
  {055010} (\bibinfo {year} {2016}{\natexlab{a}})},\ \Eprint
  {http://arxiv.org/abs/1508.04623} {arXiv:1508.04623 [hep-ph]} \BibitemShut
  {NoStop}%
\bibitem [{\citenamefont {Abada}\ \emph
  {et~al.}(2012{\natexlab{a}})\citenamefont {Abada}, \citenamefont {Das},
  \citenamefont {Vicente},\ and\ \citenamefont {Weiland}}]{Abada:2012cq}%
  \BibitemOpen
  \bibfield  {author} {\bibinfo {author} {\bibfnamefont {A.}~\bibnamefont
  {Abada}}, \bibinfo {author} {\bibfnamefont {D.}~\bibnamefont {Das}}, \bibinfo
  {author} {\bibfnamefont {A.}~\bibnamefont {Vicente}}, \ and\ \bibinfo
  {author} {\bibfnamefont {C.}~\bibnamefont {Weiland}},\ }\href {\doibase
  10.1007/JHEP09(2012)015} {\bibfield  {journal} {\bibinfo  {journal} {JHEP}\
  }\textbf {\bibinfo {volume} {09}},\ \bibinfo {pages} {015} (\bibinfo {year}
  {2012}{\natexlab{a}})},\ \Eprint {http://arxiv.org/abs/1206.6497}
  {arXiv:1206.6497 [hep-ph]} \BibitemShut {NoStop}%
\bibitem [{\citenamefont {Abada}\ \emph
  {et~al.}(2012{\natexlab{b}})\citenamefont {Abada}, \citenamefont {Das},\ and\
  \citenamefont {Weiland}}]{Abada:2011hm}%
  \BibitemOpen
  \bibfield  {author} {\bibinfo {author} {\bibfnamefont {A.}~\bibnamefont
  {Abada}}, \bibinfo {author} {\bibfnamefont {D.}~\bibnamefont {Das}}, \ and\
  \bibinfo {author} {\bibfnamefont {C.}~\bibnamefont {Weiland}},\ }\href
  {\doibase 10.1007/JHEP03(2012)100} {\bibfield  {journal} {\bibinfo  {journal}
  {JHEP}\ }\textbf {\bibinfo {volume} {03}},\ \bibinfo {pages} {100} (\bibinfo
  {year} {2012}{\natexlab{b}})},\ \Eprint {http://arxiv.org/abs/1111.5836}
  {arXiv:1111.5836 [hep-ph]} \BibitemShut {NoStop}%
\bibitem [{\citenamefont {Abada}\ \emph {et~al.}(2011)\citenamefont {Abada},
  \citenamefont {Bhattacharyya}, \citenamefont {Das},\ and\ \citenamefont
  {Weiland}}]{Abada:2010ym}%
  \BibitemOpen
  \bibfield  {author} {\bibinfo {author} {\bibfnamefont {A.}~\bibnamefont
  {Abada}}, \bibinfo {author} {\bibfnamefont {G.}~\bibnamefont
  {Bhattacharyya}}, \bibinfo {author} {\bibfnamefont {D.}~\bibnamefont {Das}},
  \ and\ \bibinfo {author} {\bibfnamefont {C.}~\bibnamefont {Weiland}},\ }\href
  {\doibase 10.1016/j.physletb.2011.05.020} {\bibfield  {journal} {\bibinfo
  {journal} {Phys. Lett.}\ }\textbf {\bibinfo {volume} {B700}},\ \bibinfo
  {pages} {351} (\bibinfo {year} {2011})},\ \Eprint
  {http://arxiv.org/abs/1011.5037} {arXiv:1011.5037 [hep-ph]} \BibitemShut
  {NoStop}%
\bibitem [{\citenamefont {Arganda}\ \emph {et~al.}(2015)\citenamefont
  {Arganda}, \citenamefont {Herrero}, \citenamefont {Marcano},\ and\
  \citenamefont {Weiland}}]{Arganda:2014dta}%
  \BibitemOpen
  \bibfield  {author} {\bibinfo {author} {\bibfnamefont {E.}~\bibnamefont
  {Arganda}}, \bibinfo {author} {\bibfnamefont {M.~J.}\ \bibnamefont
  {Herrero}}, \bibinfo {author} {\bibfnamefont {X.}~\bibnamefont {Marcano}}, \
  and\ \bibinfo {author} {\bibfnamefont {C.}~\bibnamefont {Weiland}},\ }\href
  {\doibase 10.1103/PhysRevD.91.015001} {\bibfield  {journal} {\bibinfo
  {journal} {Phys. Rev.}\ }\textbf {\bibinfo {volume} {D91}},\ \bibinfo {pages}
  {015001} (\bibinfo {year} {2015})},\ \Eprint {http://arxiv.org/abs/1405.4300}
  {arXiv:1405.4300 [hep-ph]} \BibitemShut {NoStop}%
\bibitem [{\citenamefont {Arganda}\ \emph
  {et~al.}(2016{\natexlab{b}})\citenamefont {Arganda}, \citenamefont {Herrero},
  \citenamefont {Marcano},\ and\ \citenamefont {Weiland}}]{Arganda:2015ija}%
  \BibitemOpen
  \bibfield  {author} {\bibinfo {author} {\bibfnamefont {E.}~\bibnamefont
  {Arganda}}, \bibinfo {author} {\bibfnamefont {M.~J.}\ \bibnamefont
  {Herrero}}, \bibinfo {author} {\bibfnamefont {X.}~\bibnamefont {Marcano}}, \
  and\ \bibinfo {author} {\bibfnamefont {C.}~\bibnamefont {Weiland}},\ }\href
  {\doibase 10.1016/j.physletb.2015.11.013} {\bibfield  {journal} {\bibinfo
  {journal} {Phys. Lett.}\ }\textbf {\bibinfo {volume} {B752}},\ \bibinfo
  {pages} {46} (\bibinfo {year} {2016}{\natexlab{b}})},\ \Eprint
  {http://arxiv.org/abs/1508.05074} {arXiv:1508.05074 [hep-ph]} \BibitemShut
  {NoStop}%
\bibitem [{\citenamefont {De~Romeri}\ \emph {et~al.}(2017)\citenamefont
  {De~Romeri}, \citenamefont {Herrero}, \citenamefont {Marcano},\ and\
  \citenamefont {Scarcella}}]{DeRomeri:2016gum}%
  \BibitemOpen
  \bibfield  {author} {\bibinfo {author} {\bibfnamefont {V.}~\bibnamefont
  {De~Romeri}}, \bibinfo {author} {\bibfnamefont {M.~J.}\ \bibnamefont
  {Herrero}}, \bibinfo {author} {\bibfnamefont {X.}~\bibnamefont {Marcano}}, \
  and\ \bibinfo {author} {\bibfnamefont {F.}~\bibnamefont {Scarcella}},\ }\href
  {\doibase 10.1103/PhysRevD.95.075028} {\bibfield  {journal} {\bibinfo
  {journal} {Phys. Rev.}\ }\textbf {\bibinfo {volume} {D95}},\ \bibinfo {pages}
  {075028} (\bibinfo {year} {2017})},\ \Eprint
  {http://arxiv.org/abs/1607.05257} {arXiv:1607.05257 [hep-ph]} \BibitemShut
  {NoStop}%
\bibitem [{\citenamefont {Arganda}\ \emph {et~al.}(2017)\citenamefont
  {Arganda}, \citenamefont {Herrero}, \citenamefont {Marcano}, \citenamefont
  {Morales},\ and\ \citenamefont {Szynkman}}]{Arganda:2017vdb}%
  \BibitemOpen
  \bibfield  {author} {\bibinfo {author} {\bibfnamefont {E.}~\bibnamefont
  {Arganda}}, \bibinfo {author} {\bibfnamefont {M.}~\bibnamefont {Herrero}},
  \bibinfo {author} {\bibfnamefont {X.}~\bibnamefont {Marcano}}, \bibinfo
  {author} {\bibfnamefont {R.}~\bibnamefont {Morales}}, \ and\ \bibinfo
  {author} {\bibfnamefont {A.}~\bibnamefont {Szynkman}},\ }\href {\doibase
  10.1103/PhysRevD.95.095029} {\bibfield  {journal} {\bibinfo  {journal} {Phys.
  Rev.}\ }\textbf {\bibinfo {volume} {D95}},\ \bibinfo {pages} {095029}
  (\bibinfo {year} {2017})},\ \Eprint {http://arxiv.org/abs/1612.09290}
  {arXiv:1612.09290 [hep-ph]} \BibitemShut {NoStop}%
\bibitem [{\citenamefont {Chen}\ and\ \citenamefont {Dev}(2012)}]{Chen:2011hc}%
  \BibitemOpen
  \bibfield  {author} {\bibinfo {author} {\bibfnamefont {C.-Y.}\ \bibnamefont
  {Chen}}\ and\ \bibinfo {author} {\bibfnamefont {P.~S.~B.}\ \bibnamefont
  {Dev}},\ }\href {\doibase 10.1103/PhysRevD.85.093018} {\bibfield  {journal}
  {\bibinfo  {journal} {Phys. Rev.}\ }\textbf {\bibinfo {volume} {D85}},\
  \bibinfo {pages} {093018} (\bibinfo {year} {2012})},\ \Eprint
  {http://arxiv.org/abs/1112.6419} {arXiv:1112.6419 [hep-ph]} \BibitemShut
  {NoStop}%
\bibitem [{\citenamefont {Das}\ and\ \citenamefont {Okada}(2013)}]{Das:2012ze}%
  \BibitemOpen
  \bibfield  {author} {\bibinfo {author} {\bibfnamefont {A.}~\bibnamefont
  {Das}}\ and\ \bibinfo {author} {\bibfnamefont {N.}~\bibnamefont {Okada}},\
  }\href {\doibase 10.1103/PhysRevD.88.113001} {\bibfield  {journal} {\bibinfo
  {journal} {Phys. Rev.}\ }\textbf {\bibinfo {volume} {D88}},\ \bibinfo {pages}
  {113001} (\bibinfo {year} {2013})},\ \Eprint {http://arxiv.org/abs/1207.3734}
  {arXiv:1207.3734 [hep-ph]} \BibitemShut {NoStop}%
\bibitem [{\citenamefont {Bandyopadhyay}\ \emph {et~al.}(2013)\citenamefont
  {Bandyopadhyay}, \citenamefont {Chun}, \citenamefont {Okada},\ and\
  \citenamefont {Park}}]{Bandyopadhyay:2012px}%
  \BibitemOpen
  \bibfield  {author} {\bibinfo {author} {\bibfnamefont {P.}~\bibnamefont
  {Bandyopadhyay}}, \bibinfo {author} {\bibfnamefont {E.~J.}\ \bibnamefont
  {Chun}}, \bibinfo {author} {\bibfnamefont {H.}~\bibnamefont {Okada}}, \ and\
  \bibinfo {author} {\bibfnamefont {J.-C.}\ \bibnamefont {Park}},\ }\href
  {\doibase 10.1007/JHEP01(2013)079} {\bibfield  {journal} {\bibinfo  {journal}
  {JHEP}\ }\textbf {\bibinfo {volume} {01}},\ \bibinfo {pages} {079} (\bibinfo
  {year} {2013})},\ \Eprint {http://arxiv.org/abs/1209.4803} {arXiv:1209.4803
  [hep-ph]} \BibitemShut {NoStop}%
\bibitem [{\citenamefont {Dev}\ \emph {et~al.}(2014)\citenamefont {Dev},
  \citenamefont {Pilaftsis},\ and\ \citenamefont {Yang}}]{Dev:2013wba}%
  \BibitemOpen
  \bibfield  {author} {\bibinfo {author} {\bibfnamefont {P.~S.~B.}\
  \bibnamefont {Dev}}, \bibinfo {author} {\bibfnamefont {A.}~\bibnamefont
  {Pilaftsis}}, \ and\ \bibinfo {author} {\bibfnamefont {U.-k.}\ \bibnamefont
  {Yang}},\ }\href {\doibase 10.1103/PhysRevLett.112.081801} {\bibfield
  {journal} {\bibinfo  {journal} {Phys. Rev. Lett.}\ }\textbf {\bibinfo
  {volume} {112}},\ \bibinfo {pages} {081801} (\bibinfo {year} {2014})},\
  \Eprint {http://arxiv.org/abs/1308.2209} {arXiv:1308.2209 [hep-ph]}
  \BibitemShut {NoStop}%
\bibitem [{\citenamefont {Das}\ \emph {et~al.}(2014)\citenamefont {Das},
  \citenamefont {Bhupal~Dev},\ and\ \citenamefont {Okada}}]{Das:2014jxa}%
  \BibitemOpen
  \bibfield  {author} {\bibinfo {author} {\bibfnamefont {A.}~\bibnamefont
  {Das}}, \bibinfo {author} {\bibfnamefont {P.~S.}\ \bibnamefont {Bhupal~Dev}},
  \ and\ \bibinfo {author} {\bibfnamefont {N.}~\bibnamefont {Okada}},\ }\href
  {\doibase 10.1016/j.physletb.2014.06.058} {\bibfield  {journal} {\bibinfo
  {journal} {Phys. Lett.}\ }\textbf {\bibinfo {volume} {B735}},\ \bibinfo
  {pages} {364} (\bibinfo {year} {2014})},\ \Eprint
  {http://arxiv.org/abs/1405.0177} {arXiv:1405.0177 [hep-ph]} \BibitemShut
  {NoStop}%
\bibitem [{\citenamefont {Deppisch}\ \emph {et~al.}(2015)\citenamefont
  {Deppisch}, \citenamefont {Bhupal~Dev},\ and\ \citenamefont
  {Pilaftsis}}]{Deppisch:2015qwa}%
  \BibitemOpen
  \bibfield  {author} {\bibinfo {author} {\bibfnamefont {F.~F.}\ \bibnamefont
  {Deppisch}}, \bibinfo {author} {\bibfnamefont {P.~S.}\ \bibnamefont
  {Bhupal~Dev}}, \ and\ \bibinfo {author} {\bibfnamefont {A.}~\bibnamefont
  {Pilaftsis}},\ }\href {\doibase 10.1088/1367-2630/17/7/075019} {\bibfield
  {journal} {\bibinfo  {journal} {New J. Phys.}\ }\textbf {\bibinfo {volume}
  {17}},\ \bibinfo {pages} {075019} (\bibinfo {year} {2015})},\ \Eprint
  {http://arxiv.org/abs/1502.06541} {arXiv:1502.06541 [hep-ph]} \BibitemShut
  {NoStop}%
\bibitem [{\citenamefont {Das}\ and\ \citenamefont
  {Okada}(2016)}]{Das:2015toa}%
  \BibitemOpen
  \bibfield  {author} {\bibinfo {author} {\bibfnamefont {A.}~\bibnamefont
  {Das}}\ and\ \bibinfo {author} {\bibfnamefont {N.}~\bibnamefont {Okada}},\
  }\href {\doibase 10.1103/PhysRevD.93.033003} {\bibfield  {journal} {\bibinfo
  {journal} {Phys. Rev.}\ }\textbf {\bibinfo {volume} {D93}},\ \bibinfo {pages}
  {033003} (\bibinfo {year} {2016})},\ \Eprint
  {http://arxiv.org/abs/1510.04790} {arXiv:1510.04790 [hep-ph]} \BibitemShut
  {NoStop}%
\bibitem [{\citenamefont {Mondal}\ and\ \citenamefont
  {Rai}(2016)}]{Mondal:2016kof}%
  \BibitemOpen
  \bibfield  {author} {\bibinfo {author} {\bibfnamefont {S.}~\bibnamefont
  {Mondal}}\ and\ \bibinfo {author} {\bibfnamefont {S.~K.}\ \bibnamefont
  {Rai}},\ }\href {\doibase 10.1103/PhysRevD.94.033008} {\bibfield  {journal}
  {\bibinfo  {journal} {Phys. Rev.}\ }\textbf {\bibinfo {volume} {D94}},\
  \bibinfo {pages} {033008} (\bibinfo {year} {2016})},\ \Eprint
  {http://arxiv.org/abs/1605.04508} {arXiv:1605.04508 [hep-ph]} \BibitemShut
  {NoStop}%
\bibitem [{\citenamefont {Banerjee}\ \emph {et~al.}(2015)\citenamefont
  {Banerjee}, \citenamefont {Dev}, \citenamefont {Ibarra}, \citenamefont
  {Mandal},\ and\ \citenamefont {Mitra}}]{Banerjee:2015gca}%
  \BibitemOpen
  \bibfield  {author} {\bibinfo {author} {\bibfnamefont {S.}~\bibnamefont
  {Banerjee}}, \bibinfo {author} {\bibfnamefont {P.~S.~B.}\ \bibnamefont
  {Dev}}, \bibinfo {author} {\bibfnamefont {A.}~\bibnamefont {Ibarra}},
  \bibinfo {author} {\bibfnamefont {T.}~\bibnamefont {Mandal}}, \ and\ \bibinfo
  {author} {\bibfnamefont {M.}~\bibnamefont {Mitra}},\ }\href {\doibase
  10.1103/PhysRevD.92.075002} {\bibfield  {journal} {\bibinfo  {journal} {Phys.
  Rev.}\ }\textbf {\bibinfo {volume} {D92}},\ \bibinfo {pages} {075002}
  (\bibinfo {year} {2015})},\ \Eprint {http://arxiv.org/abs/1503.05491}
  {arXiv:1503.05491 [hep-ph]} \BibitemShut {NoStop}%
\bibitem [{\citenamefont {Das}(2017)}]{Das:2017pvt}%
  \BibitemOpen
  \bibfield  {author} {\bibinfo {author} {\bibfnamefont {A.}~\bibnamefont
  {Das}},\ }\href@noop {} {\  (\bibinfo {year} {2017})},\ \Eprint
  {http://arxiv.org/abs/1701.04946} {arXiv:1701.04946 [hep-ph]} \BibitemShut
  {NoStop}%
\bibitem [{\citenamefont {Alva}\ \emph {et~al.}(2015)\citenamefont {Alva},
  \citenamefont {Han},\ and\ \citenamefont {Ruiz}}]{Alva:2014gxa}%
  \BibitemOpen
  \bibfield  {author} {\bibinfo {author} {\bibfnamefont {D.}~\bibnamefont
  {Alva}}, \bibinfo {author} {\bibfnamefont {T.}~\bibnamefont {Han}}, \ and\
  \bibinfo {author} {\bibfnamefont {R.}~\bibnamefont {Ruiz}},\ }\href {\doibase
  10.1007/JHEP02(2015)072} {\bibfield  {journal} {\bibinfo  {journal} {JHEP}\
  }\textbf {\bibinfo {volume} {02}},\ \bibinfo {pages} {072} (\bibinfo {year}
  {2015})},\ \Eprint {http://arxiv.org/abs/1411.7305} {arXiv:1411.7305
  [hep-ph]} \BibitemShut {NoStop}%
\bibitem [{\citenamefont {Degrande}\ \emph {et~al.}(2016)\citenamefont
  {Degrande}, \citenamefont {Mattelaer}, \citenamefont {Ruiz},\ and\
  \citenamefont {Turner}}]{Degrande:2016aje}%
  \BibitemOpen
  \bibfield  {author} {\bibinfo {author} {\bibfnamefont {C.}~\bibnamefont
  {Degrande}}, \bibinfo {author} {\bibfnamefont {O.}~\bibnamefont {Mattelaer}},
  \bibinfo {author} {\bibfnamefont {R.}~\bibnamefont {Ruiz}}, \ and\ \bibinfo
  {author} {\bibfnamefont {J.}~\bibnamefont {Turner}},\ }\href {\doibase
  10.1103/PhysRevD.94.053002} {\bibfield  {journal} {\bibinfo  {journal} {Phys.
  Rev.}\ }\textbf {\bibinfo {volume} {D94}},\ \bibinfo {pages} {053002}
  (\bibinfo {year} {2016})},\ \Eprint {http://arxiv.org/abs/1602.06957}
  {arXiv:1602.06957 [hep-ph]} \BibitemShut {NoStop}%
\bibitem [{\citenamefont {Doršner}\ \emph {et~al.}(2016)\citenamefont
  {Doršner}, \citenamefont {Fajfer}, \citenamefont {Greljo}, \citenamefont
  {Kamenik},\ and\ \citenamefont {Košnik}}]{Dorsner:2016wpm}%
  \BibitemOpen
  \bibfield  {author} {\bibinfo {author} {\bibfnamefont {I.}~\bibnamefont
  {Doršner}}, \bibinfo {author} {\bibfnamefont {S.}~\bibnamefont {Fajfer}},
  \bibinfo {author} {\bibfnamefont {A.}~\bibnamefont {Greljo}}, \bibinfo
  {author} {\bibfnamefont {J.~F.}\ \bibnamefont {Kamenik}}, \ and\ \bibinfo
  {author} {\bibfnamefont {N.}~\bibnamefont {Košnik}},\ }\href {\doibase
  10.1016/j.physrep.2016.06.001} {\bibfield  {journal} {\bibinfo  {journal}
  {Phys. Rept.}\ }\textbf {\bibinfo {volume} {641}},\ \bibinfo {pages} {1}
  (\bibinfo {year} {2016})},\ \Eprint {http://arxiv.org/abs/1603.04993}
  {arXiv:1603.04993 [hep-ph]} \BibitemShut {NoStop}%
\bibitem [{\citenamefont {Pati}\ and\ \citenamefont
  {Salam}(1974)}]{Pati:1974yy}%
  \BibitemOpen
  \bibfield  {author} {\bibinfo {author} {\bibfnamefont {J.~C.}\ \bibnamefont
  {Pati}}\ and\ \bibinfo {author} {\bibfnamefont {A.}~\bibnamefont {Salam}},\
  }\href {\doibase 10.1103/PhysRevD.10.275, 10.1103/PhysRevD.11.703.2}
  {\bibfield  {journal} {\bibinfo  {journal} {Phys. Rev.}\ }\textbf {\bibinfo
  {volume} {D10}},\ \bibinfo {pages} {275} (\bibinfo {year} {1974})},\ \bibinfo
  {note} {[Erratum: Phys. Rev.D11,703(1975)]}\BibitemShut {NoStop}%
\bibitem [{\citenamefont {Georgi}\ and\ \citenamefont
  {Glashow}(1974)}]{Georgi:1974sy}%
  \BibitemOpen
  \bibfield  {author} {\bibinfo {author} {\bibfnamefont {H.}~\bibnamefont
  {Georgi}}\ and\ \bibinfo {author} {\bibfnamefont {S.~L.}\ \bibnamefont
  {Glashow}},\ }\href {\doibase 10.1103/PhysRevLett.32.438} {\bibfield
  {journal} {\bibinfo  {journal} {Phys. Rev. Lett.}\ }\textbf {\bibinfo
  {volume} {32}},\ \bibinfo {pages} {438} (\bibinfo {year} {1974})}\BibitemShut
  {NoStop}%
\bibitem [{\citenamefont {Georgi}(1975)}]{Georgi:1974my}%
  \BibitemOpen
  \bibfield  {author} {\bibinfo {author} {\bibfnamefont {H.}~\bibnamefont
  {Georgi}},\ }\bibfield  {booktitle} {\emph {\bibinfo {booktitle} {{PARTICLES
  AND FIELDS — 1974: Proceedings of the Williamsburg Meeting of APS/DPF}}},\
  }\href {\doibase 10.1063/1.2947450} {\bibfield  {journal} {\bibinfo
  {journal} {AIP Conf. Proc.}\ }\textbf {\bibinfo {volume} {23}},\ \bibinfo
  {pages} {575} (\bibinfo {year} {1975})}\BibitemShut {NoStop}%
\bibitem [{\citenamefont {Fritzsch}\ and\ \citenamefont
  {Minkowski}(1975)}]{Fritzsch:1974nn}%
  \BibitemOpen
  \bibfield  {author} {\bibinfo {author} {\bibfnamefont {H.}~\bibnamefont
  {Fritzsch}}\ and\ \bibinfo {author} {\bibfnamefont {P.}~\bibnamefont
  {Minkowski}},\ }\href {\doibase 10.1016/0003-4916(75)90211-0} {\bibfield
  {journal} {\bibinfo  {journal} {Annals Phys.}\ }\textbf {\bibinfo {volume}
  {93}},\ \bibinfo {pages} {193} (\bibinfo {year} {1975})}\BibitemShut
  {NoStop}%
\bibitem [{\citenamefont {Altmannshofer}\ \emph {et~al.}(2016)\citenamefont
  {Altmannshofer}, \citenamefont {Carena},\ and\ \citenamefont
  {Crivellin}}]{Altmannshofer:2016oaq}%
  \BibitemOpen
  \bibfield  {author} {\bibinfo {author} {\bibfnamefont {W.}~\bibnamefont
  {Altmannshofer}}, \bibinfo {author} {\bibfnamefont {M.}~\bibnamefont
  {Carena}}, \ and\ \bibinfo {author} {\bibfnamefont {A.}~\bibnamefont
  {Crivellin}},\ }\href {\doibase 10.1103/PhysRevD.94.095026} {\bibfield
  {journal} {\bibinfo  {journal} {Phys. Rev.}\ }\textbf {\bibinfo {volume}
  {D94}},\ \bibinfo {pages} {095026} (\bibinfo {year} {2016})},\ \Eprint
  {http://arxiv.org/abs/1604.08221} {arXiv:1604.08221 [hep-ph]} \BibitemShut
  {NoStop}%
\bibitem [{\citenamefont {Bečirević}\ \emph {et~al.}(2016)\citenamefont
  {Bečirević}, \citenamefont {Fajfer}, \citenamefont {Košnik},\ and\
  \citenamefont {Sumensari}}]{Becirevic:2016yqi}%
  \BibitemOpen
  \bibfield  {author} {\bibinfo {author} {\bibfnamefont {D.}~\bibnamefont
  {Bečirević}}, \bibinfo {author} {\bibfnamefont {S.}~\bibnamefont {Fajfer}},
  \bibinfo {author} {\bibfnamefont {N.}~\bibnamefont {Košnik}}, \ and\
  \bibinfo {author} {\bibfnamefont {O.}~\bibnamefont {Sumensari}},\ }\href
  {\doibase 10.1103/PhysRevD.94.115021} {\bibfield  {journal} {\bibinfo
  {journal} {Phys. Rev.}\ }\textbf {\bibinfo {volume} {D94}},\ \bibinfo {pages}
  {115021} (\bibinfo {year} {2016})},\ \Eprint
  {http://arxiv.org/abs/1608.08501} {arXiv:1608.08501 [hep-ph]} \BibitemShut
  {NoStop}%
\bibitem [{\citenamefont {Bečirević}\ \emph {et~al.}(2015)\citenamefont
  {Bečirević}, \citenamefont {Fajfer},\ and\ \citenamefont
  {Košnik}}]{Becirevic:2015asa}%
  \BibitemOpen
  \bibfield  {author} {\bibinfo {author} {\bibfnamefont {D.}~\bibnamefont
  {Bečirević}}, \bibinfo {author} {\bibfnamefont {S.}~\bibnamefont {Fajfer}},
  \ and\ \bibinfo {author} {\bibfnamefont {N.}~\bibnamefont {Košnik}},\ }\href
  {\doibase 10.1103/PhysRevD.92.014016} {\bibfield  {journal} {\bibinfo
  {journal} {Phys. Rev.}\ }\textbf {\bibinfo {volume} {D92}},\ \bibinfo {pages}
  {014016} (\bibinfo {year} {2015})},\ \Eprint
  {http://arxiv.org/abs/1503.09024} {arXiv:1503.09024 [hep-ph]} \BibitemShut
  {NoStop}%
\bibitem [{\citenamefont {Doršner}\ \emph {et~al.}(2013)\citenamefont
  {Doršner}, \citenamefont {Fajfer}, \citenamefont {Košnik},\ and\
  \citenamefont {Nišandžić}}]{Dorsner:2013tla}%
  \BibitemOpen
  \bibfield  {author} {\bibinfo {author} {\bibfnamefont {I.}~\bibnamefont
  {Doršner}}, \bibinfo {author} {\bibfnamefont {S.}~\bibnamefont {Fajfer}},
  \bibinfo {author} {\bibfnamefont {N.}~\bibnamefont {Košnik}}, \ and\
  \bibinfo {author} {\bibfnamefont {I.}~\bibnamefont {Nišandžić}},\ }\href
  {\doibase 10.1007/JHEP11(2013)084} {\bibfield  {journal} {\bibinfo  {journal}
  {JHEP}\ }\textbf {\bibinfo {volume} {11}},\ \bibinfo {pages} {084} (\bibinfo
  {year} {2013})},\ \Eprint {http://arxiv.org/abs/1306.6493} {arXiv:1306.6493
  [hep-ph]} \BibitemShut {NoStop}%
\bibitem [{\citenamefont {Abada}\ and\ \citenamefont
  {Lucente}(2014)}]{Abada:2014vea}%
  \BibitemOpen
  \bibfield  {author} {\bibinfo {author} {\bibfnamefont {A.}~\bibnamefont
  {Abada}}\ and\ \bibinfo {author} {\bibfnamefont {M.}~\bibnamefont
  {Lucente}},\ }\href {\doibase 10.1016/j.nuclphysb.2014.06.003} {\bibfield
  {journal} {\bibinfo  {journal} {Nucl. Phys.}\ }\textbf {\bibinfo {volume}
  {B885}},\ \bibinfo {pages} {651} (\bibinfo {year} {2014})},\ \Eprint
  {http://arxiv.org/abs/1401.1507} {arXiv:1401.1507 [hep-ph]} \BibitemShut
  {NoStop}%
\bibitem [{\citenamefont {with Olcyr~Sumensari}()}]{pc:olcyr}%
  \BibitemOpen
  \bibfield  {author} {\bibinfo {author} {\bibfnamefont {P.~C.}\ \bibnamefont
  {with Olcyr~Sumensari}},\ }\href@noop {} {\ }\BibitemShut {NoStop}%
\bibitem [{\citenamefont {Christensen}\ and\ \citenamefont
  {Duhr}(2009)}]{Christensen:2008py}%
  \BibitemOpen
  \bibfield  {author} {\bibinfo {author} {\bibfnamefont {N.~D.}\ \bibnamefont
  {Christensen}}\ and\ \bibinfo {author} {\bibfnamefont {C.}~\bibnamefont
  {Duhr}},\ }\href {\doibase 10.1016/j.cpc.2009.02.018} {\bibfield  {journal}
  {\bibinfo  {journal} {Comput. Phys. Commun.}\ }\textbf {\bibinfo {volume}
  {180}},\ \bibinfo {pages} {1614} (\bibinfo {year} {2009})},\ \Eprint
  {http://arxiv.org/abs/0806.4194} {arXiv:0806.4194 [hep-ph]} \BibitemShut
  {NoStop}%
\bibitem [{\citenamefont {Alloul}\ \emph {et~al.}(2014)\citenamefont {Alloul},
  \citenamefont {Christensen}, \citenamefont {Degrande}, \citenamefont {Duhr},\
  and\ \citenamefont {Fuks}}]{Alloul:2013bka}%
  \BibitemOpen
  \bibfield  {author} {\bibinfo {author} {\bibfnamefont {A.}~\bibnamefont
  {Alloul}}, \bibinfo {author} {\bibfnamefont {N.~D.}\ \bibnamefont
  {Christensen}}, \bibinfo {author} {\bibfnamefont {C.}~\bibnamefont
  {Degrande}}, \bibinfo {author} {\bibfnamefont {C.}~\bibnamefont {Duhr}}, \
  and\ \bibinfo {author} {\bibfnamefont {B.}~\bibnamefont {Fuks}},\ }\href
  {\doibase 10.1016/j.cpc.2014.04.012} {\bibfield  {journal} {\bibinfo
  {journal} {Comput. Phys. Commun.}\ }\textbf {\bibinfo {volume} {185}},\
  \bibinfo {pages} {2250} (\bibinfo {year} {2014})},\ \Eprint
  {http://arxiv.org/abs/1310.1921} {arXiv:1310.1921 [hep-ph]} \BibitemShut
  {NoStop}%
\bibitem [{\citenamefont {Alwall}\ \emph {et~al.}(2011)\citenamefont {Alwall},
  \citenamefont {Herquet}, \citenamefont {Maltoni}, \citenamefont {Mattelaer},\
  and\ \citenamefont {Stelzer}}]{Alwall:2011uj}%
  \BibitemOpen
  \bibfield  {author} {\bibinfo {author} {\bibfnamefont {J.}~\bibnamefont
  {Alwall}}, \bibinfo {author} {\bibfnamefont {M.}~\bibnamefont {Herquet}},
  \bibinfo {author} {\bibfnamefont {F.}~\bibnamefont {Maltoni}}, \bibinfo
  {author} {\bibfnamefont {O.}~\bibnamefont {Mattelaer}}, \ and\ \bibinfo
  {author} {\bibfnamefont {T.}~\bibnamefont {Stelzer}},\ }\href {\doibase
  10.1007/JHEP06(2011)128} {\bibfield  {journal} {\bibinfo  {journal} {JHEP}\
  }\textbf {\bibinfo {volume} {06}},\ \bibinfo {pages} {128} (\bibinfo {year}
  {2011})},\ \Eprint {http://arxiv.org/abs/1106.0522} {arXiv:1106.0522
  [hep-ph]} \BibitemShut {NoStop}%
\bibitem [{\citenamefont {Alwall}\ \emph {et~al.}(2014)\citenamefont {Alwall},
  \citenamefont {Frederix}, \citenamefont {Frixione}, \citenamefont {Hirschi},
  \citenamefont {Maltoni}, \citenamefont {Mattelaer}, \citenamefont {Shao},
  \citenamefont {Stelzer}, \citenamefont {Torrielli},\ and\ \citenamefont
  {Zaro}}]{Alwall:2014hca}%
  \BibitemOpen
  \bibfield  {author} {\bibinfo {author} {\bibfnamefont {J.}~\bibnamefont
  {Alwall}}, \bibinfo {author} {\bibfnamefont {R.}~\bibnamefont {Frederix}},
  \bibinfo {author} {\bibfnamefont {S.}~\bibnamefont {Frixione}}, \bibinfo
  {author} {\bibfnamefont {V.}~\bibnamefont {Hirschi}}, \bibinfo {author}
  {\bibfnamefont {F.}~\bibnamefont {Maltoni}}, \bibinfo {author} {\bibfnamefont
  {O.}~\bibnamefont {Mattelaer}}, \bibinfo {author} {\bibfnamefont {H.~S.}\
  \bibnamefont {Shao}}, \bibinfo {author} {\bibfnamefont {T.}~\bibnamefont
  {Stelzer}}, \bibinfo {author} {\bibfnamefont {P.}~\bibnamefont {Torrielli}},
  \ and\ \bibinfo {author} {\bibfnamefont {M.}~\bibnamefont {Zaro}},\ }\href
  {\doibase 10.1007/JHEP07(2014)079} {\bibfield  {journal} {\bibinfo  {journal}
  {JHEP}\ }\textbf {\bibinfo {volume} {07}},\ \bibinfo {pages} {079} (\bibinfo
  {year} {2014})},\ \Eprint {http://arxiv.org/abs/1405.0301} {arXiv:1405.0301
  [hep-ph]} \BibitemShut {NoStop}%
\bibitem [{\citenamefont {Ball}\ \emph {et~al.}(2013)\citenamefont {Ball} \emph
  {et~al.}}]{Ball:2012cx}%
  \BibitemOpen
  \bibfield  {author} {\bibinfo {author} {\bibfnamefont {R.~D.}\ \bibnamefont
  {Ball}} \emph {et~al.},\ }\href {\doibase 10.1016/j.nuclphysb.2012.10.003}
  {\bibfield  {journal} {\bibinfo  {journal} {Nucl. Phys.}\ }\textbf {\bibinfo
  {volume} {B867}},\ \bibinfo {pages} {244} (\bibinfo {year} {2013})},\ \Eprint
  {http://arxiv.org/abs/1207.1303} {arXiv:1207.1303 [hep-ph]} \BibitemShut
  {NoStop}%
\bibitem [{\citenamefont {Ball}\ \emph {et~al.}(2015)\citenamefont {Ball} \emph
  {et~al.}}]{Ball:2014uwa}%
  \BibitemOpen
  \bibfield  {author} {\bibinfo {author} {\bibfnamefont {R.~D.}\ \bibnamefont
  {Ball}} \emph {et~al.} (\bibinfo {collaboration} {NNPDF}),\ }\href {\doibase
  10.1007/JHEP04(2015)040} {\bibfield  {journal} {\bibinfo  {journal} {JHEP}\
  }\textbf {\bibinfo {volume} {04}},\ \bibinfo {pages} {040} (\bibinfo {year}
  {2015})},\ \Eprint {http://arxiv.org/abs/1410.8849} {arXiv:1410.8849
  [hep-ph]} \BibitemShut {NoStop}%
\bibitem [{\citenamefont
  {{\url{https://cp3.irmp.ucl.ac.be/projects/madgraph/wiki/FAQ-General-13}}}()}]{madgraph_scale}%
  \BibitemOpen
  \bibfield  {author} {\bibinfo {author} {\bibnamefont
  {{\url{https://cp3.irmp.ucl.ac.be/projects/madgraph/wiki/FAQ-General-13}}}},\
  }\href@noop {} {\ }\BibitemShut {NoStop}%
\bibitem [{\citenamefont {Diaz}\ \emph {et~al.}(2017)\citenamefont {Diaz},
  \citenamefont {Schmaltz},\ and\ \citenamefont {Zhong}}]{Diaz:2017lit}%
  \BibitemOpen
  \bibfield  {author} {\bibinfo {author} {\bibfnamefont {B.}~\bibnamefont
  {Diaz}}, \bibinfo {author} {\bibfnamefont {M.}~\bibnamefont {Schmaltz}}, \
  and\ \bibinfo {author} {\bibfnamefont {Y.-M.}\ \bibnamefont {Zhong}},\ }\href
  {\doibase 10.1007/JHEP10(2017)097} {\bibfield  {journal} {\bibinfo  {journal}
  {JHEP}\ }\textbf {\bibinfo {volume} {10}},\ \bibinfo {pages} {097} (\bibinfo
  {year} {2017})},\ \Eprint {http://arxiv.org/abs/1706.05033} {arXiv:1706.05033
  [hep-ph]} \BibitemShut {NoStop}%
\bibitem [{\citenamefont {Mandal}\ \emph {et~al.}(2016)\citenamefont {Mandal},
  \citenamefont {Mitra},\ and\ \citenamefont {Seth}}]{Mandal:2015lca}%
  \BibitemOpen
  \bibfield  {author} {\bibinfo {author} {\bibfnamefont {T.}~\bibnamefont
  {Mandal}}, \bibinfo {author} {\bibfnamefont {S.}~\bibnamefont {Mitra}}, \
  and\ \bibinfo {author} {\bibfnamefont {S.}~\bibnamefont {Seth}},\ }\href
  {\doibase 10.1103/PhysRevD.93.035018} {\bibfield  {journal} {\bibinfo
  {journal} {Phys. Rev.}\ }\textbf {\bibinfo {volume} {D93}},\ \bibinfo {pages}
  {035018} (\bibinfo {year} {2016})},\ \Eprint
  {http://arxiv.org/abs/1506.07369} {arXiv:1506.07369 [hep-ph]} \BibitemShut
  {NoStop}%
\bibitem [{\citenamefont {Collaboration}(2016)}]{CMS:2016hsa}%
  \BibitemOpen
  \bibfield  {author} {\bibinfo {author} {\bibfnamefont {C.}~\bibnamefont
  {Collaboration}} (\bibinfo {collaboration} {CMS}),\ }\href@noop {} {\
  (\bibinfo {year} {2016})}\BibitemShut {NoStop}%
\bibitem [{\citenamefont {Keung}\ and\ \citenamefont
  {Senjanovic}(1983)}]{Keung:1983uu}%
  \BibitemOpen
  \bibfield  {author} {\bibinfo {author} {\bibfnamefont {W.-Y.}\ \bibnamefont
  {Keung}}\ and\ \bibinfo {author} {\bibfnamefont {G.}~\bibnamefont
  {Senjanovic}},\ }\href {\doibase 10.1103/PhysRevLett.50.1427} {\bibfield
  {journal} {\bibinfo  {journal} {Phys. Rev. Lett.}\ }\textbf {\bibinfo
  {volume} {50}},\ \bibinfo {pages} {1427} (\bibinfo {year}
  {1983})}\BibitemShut {NoStop}%
\bibitem [{\citenamefont {Datta}\ \emph {et~al.}(1994)\citenamefont {Datta},
  \citenamefont {Guchait},\ and\ \citenamefont {Pilaftsis}}]{Datta:1993nm}%
  \BibitemOpen
  \bibfield  {author} {\bibinfo {author} {\bibfnamefont {A.}~\bibnamefont
  {Datta}}, \bibinfo {author} {\bibfnamefont {M.}~\bibnamefont {Guchait}}, \
  and\ \bibinfo {author} {\bibfnamefont {A.}~\bibnamefont {Pilaftsis}},\ }\href
  {\doibase 10.1103/PhysRevD.50.3195} {\bibfield  {journal} {\bibinfo
  {journal} {Phys. Rev.}\ }\textbf {\bibinfo {volume} {D50}},\ \bibinfo {pages}
  {3195} (\bibinfo {year} {1994})},\ \Eprint
  {http://arxiv.org/abs/hep-ph/9311257} {arXiv:hep-ph/9311257 [hep-ph]}
  \BibitemShut {NoStop}%
\bibitem [{\citenamefont {Almeida}\ \emph {et~al.}(2000)\citenamefont
  {Almeida}, \citenamefont {do~Amaral~Coutinho}, \citenamefont
  {Martins~Simoes},\ and\ \citenamefont {do~Vale}}]{Almeida:2000pz}%
  \BibitemOpen
  \bibfield  {author} {\bibinfo {author} {\bibfnamefont {F.~M.~L.}\
  \bibnamefont {Almeida}, \bibfnamefont {Jr.}}, \bibinfo {author}
  {\bibfnamefont {Y.}~\bibnamefont {do~Amaral~Coutinho}}, \bibinfo {author}
  {\bibfnamefont {J.~A.}\ \bibnamefont {Martins~Simoes}}, \ and\ \bibinfo
  {author} {\bibfnamefont {M.~A.~B.}\ \bibnamefont {do~Vale}},\ }\href
  {\doibase 10.1103/PhysRevD.62.075004} {\bibfield  {journal} {\bibinfo
  {journal} {Phys. Rev.}\ }\textbf {\bibinfo {volume} {D62}},\ \bibinfo {pages}
  {075004} (\bibinfo {year} {2000})},\ \Eprint
  {http://arxiv.org/abs/hep-ph/0002024} {arXiv:hep-ph/0002024 [hep-ph]}
  \BibitemShut {NoStop}%
\bibitem [{\citenamefont {Panella}\ \emph {et~al.}(2002)\citenamefont
  {Panella}, \citenamefont {Cannoni}, \citenamefont {Carimalo},\ and\
  \citenamefont {Srivastava}}]{Panella:2001wq}%
  \BibitemOpen
  \bibfield  {author} {\bibinfo {author} {\bibfnamefont {O.}~\bibnamefont
  {Panella}}, \bibinfo {author} {\bibfnamefont {M.}~\bibnamefont {Cannoni}},
  \bibinfo {author} {\bibfnamefont {C.}~\bibnamefont {Carimalo}}, \ and\
  \bibinfo {author} {\bibfnamefont {Y.~N.}\ \bibnamefont {Srivastava}},\ }\href
  {\doibase 10.1103/PhysRevD.65.035005} {\bibfield  {journal} {\bibinfo
  {journal} {Phys. Rev.}\ }\textbf {\bibinfo {volume} {D65}},\ \bibinfo {pages}
  {035005} (\bibinfo {year} {2002})},\ \Eprint
  {http://arxiv.org/abs/hep-ph/0107308} {arXiv:hep-ph/0107308 [hep-ph]}
  \BibitemShut {NoStop}%
\bibitem [{\citenamefont {Han}\ and\ \citenamefont {Zhang}(2006)}]{Han:2006ip}%
  \BibitemOpen
  \bibfield  {author} {\bibinfo {author} {\bibfnamefont {T.}~\bibnamefont
  {Han}}\ and\ \bibinfo {author} {\bibfnamefont {B.}~\bibnamefont {Zhang}},\
  }\href {\doibase 10.1103/PhysRevLett.97.171804} {\bibfield  {journal}
  {\bibinfo  {journal} {Phys. Rev. Lett.}\ }\textbf {\bibinfo {volume} {97}},\
  \bibinfo {pages} {171804} (\bibinfo {year} {2006})},\ \Eprint
  {http://arxiv.org/abs/hep-ph/0604064} {arXiv:hep-ph/0604064 [hep-ph]}
  \BibitemShut {NoStop}%
\bibitem [{\citenamefont {del Aguila}\ \emph {et~al.}(2007)\citenamefont {del
  Aguila}, \citenamefont {Aguilar-Saavedra},\ and\ \citenamefont
  {Pittau}}]{delAguila:2007qnc}%
  \BibitemOpen
  \bibfield  {author} {\bibinfo {author} {\bibfnamefont {F.}~\bibnamefont {del
  Aguila}}, \bibinfo {author} {\bibfnamefont {J.~A.}\ \bibnamefont
  {Aguilar-Saavedra}}, \ and\ \bibinfo {author} {\bibfnamefont
  {R.}~\bibnamefont {Pittau}},\ }\href {\doibase 10.1088/1126-6708/2007/10/047}
  {\bibfield  {journal} {\bibinfo  {journal} {JHEP}\ }\textbf {\bibinfo
  {volume} {10}},\ \bibinfo {pages} {047} (\bibinfo {year} {2007})},\ \Eprint
  {http://arxiv.org/abs/hep-ph/0703261} {arXiv:hep-ph/0703261 [hep-ph]}
  \BibitemShut {NoStop}%
\bibitem [{\citenamefont {Huitu}\ \emph {et~al.}(2008)\citenamefont {Huitu},
  \citenamefont {Khalil}, \citenamefont {Okada},\ and\ \citenamefont
  {Rai}}]{Huitu:2008gf}%
  \BibitemOpen
  \bibfield  {author} {\bibinfo {author} {\bibfnamefont {K.}~\bibnamefont
  {Huitu}}, \bibinfo {author} {\bibfnamefont {S.}~\bibnamefont {Khalil}},
  \bibinfo {author} {\bibfnamefont {H.}~\bibnamefont {Okada}}, \ and\ \bibinfo
  {author} {\bibfnamefont {S.~K.}\ \bibnamefont {Rai}},\ }\href {\doibase
  10.1103/PhysRevLett.101.181802} {\bibfield  {journal} {\bibinfo  {journal}
  {Phys. Rev. Lett.}\ }\textbf {\bibinfo {volume} {101}},\ \bibinfo {pages}
  {181802} (\bibinfo {year} {2008})},\ \Eprint {http://arxiv.org/abs/0803.2799}
  {arXiv:0803.2799 [hep-ph]} \BibitemShut {NoStop}%
\bibitem [{\citenamefont {Atre}\ \emph {et~al.}(2009)\citenamefont {Atre},
  \citenamefont {Han}, \citenamefont {Pascoli},\ and\ \citenamefont
  {Zhang}}]{Atre:2009rg}%
  \BibitemOpen
  \bibfield  {author} {\bibinfo {author} {\bibfnamefont {A.}~\bibnamefont
  {Atre}}, \bibinfo {author} {\bibfnamefont {T.}~\bibnamefont {Han}}, \bibinfo
  {author} {\bibfnamefont {S.}~\bibnamefont {Pascoli}}, \ and\ \bibinfo
  {author} {\bibfnamefont {B.}~\bibnamefont {Zhang}},\ }\href {\doibase
  10.1088/1126-6708/2009/05/030} {\bibfield  {journal} {\bibinfo  {journal}
  {JHEP}\ }\textbf {\bibinfo {volume} {05}},\ \bibinfo {pages} {030} (\bibinfo
  {year} {2009})},\ \Eprint {http://arxiv.org/abs/0901.3589} {arXiv:0901.3589
  [hep-ph]} \BibitemShut {NoStop}%
\bibitem [{\citenamefont {Das}\ \emph {et~al.}(2016)\citenamefont {Das},
  \citenamefont {Konar},\ and\ \citenamefont {Majhi}}]{Das:2016hof}%
  \BibitemOpen
  \bibfield  {author} {\bibinfo {author} {\bibfnamefont {A.}~\bibnamefont
  {Das}}, \bibinfo {author} {\bibfnamefont {P.}~\bibnamefont {Konar}}, \ and\
  \bibinfo {author} {\bibfnamefont {S.}~\bibnamefont {Majhi}},\ }\href
  {\doibase 10.1007/JHEP06(2016)019} {\bibfield  {journal} {\bibinfo  {journal}
  {JHEP}\ }\textbf {\bibinfo {volume} {06}},\ \bibinfo {pages} {019} (\bibinfo
  {year} {2016})},\ \Eprint {http://arxiv.org/abs/1604.00608} {arXiv:1604.00608
  [hep-ph]} \BibitemShut {NoStop}%
\bibitem [{\citenamefont {Ruiz}\ \emph {et~al.}(2017)\citenamefont {Ruiz},
  \citenamefont {Spannowsky},\ and\ \citenamefont {Waite}}]{Ruiz:2017yyf}%
  \BibitemOpen
  \bibfield  {author} {\bibinfo {author} {\bibfnamefont {R.}~\bibnamefont
  {Ruiz}}, \bibinfo {author} {\bibfnamefont {M.}~\bibnamefont {Spannowsky}}, \
  and\ \bibinfo {author} {\bibfnamefont {P.}~\bibnamefont {Waite}},\
  }\href@noop {} {\  (\bibinfo {year} {2017})},\ \Eprint
  {http://arxiv.org/abs/1706.02298} {arXiv:1706.02298 [hep-ph]} \BibitemShut
  {NoStop}%
\bibitem [{\citenamefont {Abulencia}\ \emph {et~al.}(2007)\citenamefont
  {Abulencia} \emph {et~al.}}]{Abulencia:2007rd}%
  \BibitemOpen
  \bibfield  {author} {\bibinfo {author} {\bibfnamefont {A.}~\bibnamefont
  {Abulencia}} \emph {et~al.} (\bibinfo {collaboration} {CDF}),\ }\href
  {\doibase 10.1103/PhysRevLett.98.221803} {\bibfield  {journal} {\bibinfo
  {journal} {Phys. Rev. Lett.}\ }\textbf {\bibinfo {volume} {98}},\ \bibinfo
  {pages} {221803} (\bibinfo {year} {2007})},\ \Eprint
  {http://arxiv.org/abs/hep-ex/0702051} {arXiv:hep-ex/0702051 [hep-ex]}
  \BibitemShut {NoStop}%
\bibitem [{\citenamefont {Chatrchyan}\ \emph {et~al.}(2012)\citenamefont
  {Chatrchyan} \emph {et~al.}}]{Chatrchyan:2012fla}%
  \BibitemOpen
  \bibfield  {author} {\bibinfo {author} {\bibfnamefont {S.}~\bibnamefont
  {Chatrchyan}} \emph {et~al.} (\bibinfo {collaboration} {CMS}),\ }\href
  {\doibase 10.1016/j.physletb.2012.09.012} {\bibfield  {journal} {\bibinfo
  {journal} {Phys. Lett.}\ }\textbf {\bibinfo {volume} {B717}},\ \bibinfo
  {pages} {109} (\bibinfo {year} {2012})},\ \Eprint
  {http://arxiv.org/abs/1207.6079} {arXiv:1207.6079 [hep-ex]} \BibitemShut
  {NoStop}%
\bibitem [{ATL(2012)}]{ATLAS:2012yoa}%
  \BibitemOpen
  \href
  {https://atlas.web.cern.ch/Atlas/GROUPS/PHYSICS/CONFNOTES/ATLAS-CONF-2012-139/}
  {\emph {\bibinfo {title} {{Search for Majorana neutrino production in pp
  collisions at sqrt(s)=7 TeV in dimuon final states with the ATLAS
  detector}}}},\ \bibinfo {type} {Tech. Rep.}\ \bibinfo {number}
  {ATLAS-CONF-2012-139}\ (\bibinfo  {institution} {CERN},\ \bibinfo {address}
  {Geneva},\ \bibinfo {year} {2012})\BibitemShut {NoStop}%
\bibitem [{\citenamefont {Khachatryan}\ \emph {et~al.}(2015)\citenamefont
  {Khachatryan} \emph {et~al.}}]{Khachatryan:2015gha}%
  \BibitemOpen
  \bibfield  {author} {\bibinfo {author} {\bibfnamefont {V.}~\bibnamefont
  {Khachatryan}} \emph {et~al.} (\bibinfo {collaboration} {CMS}),\ }\href
  {\doibase 10.1016/j.physletb.2015.06.070} {\bibfield  {journal} {\bibinfo
  {journal} {Phys. Lett.}\ }\textbf {\bibinfo {volume} {B748}},\ \bibinfo
  {pages} {144} (\bibinfo {year} {2015})},\ \Eprint
  {http://arxiv.org/abs/1501.05566} {arXiv:1501.05566 [hep-ex]} \BibitemShut
  {NoStop}%
\bibitem [{\citenamefont {Aad}\ \emph {et~al.}(2015)\citenamefont {Aad} \emph
  {et~al.}}]{Aad:2015xaa}%
  \BibitemOpen
  \bibfield  {author} {\bibinfo {author} {\bibfnamefont {G.}~\bibnamefont
  {Aad}} \emph {et~al.} (\bibinfo {collaboration} {ATLAS}),\ }\href {\doibase
  10.1007/JHEP07(2015)162} {\bibfield  {journal} {\bibinfo  {journal} {JHEP}\
  }\textbf {\bibinfo {volume} {07}},\ \bibinfo {pages} {162} (\bibinfo {year}
  {2015})},\ \Eprint {http://arxiv.org/abs/1506.06020} {arXiv:1506.06020
  [hep-ex]} \BibitemShut {NoStop}%
\bibitem [{\citenamefont {Khachatryan}\ \emph {et~al.}(2016)\citenamefont
  {Khachatryan} \emph {et~al.}}]{Khachatryan:2016olu}%
  \BibitemOpen
  \bibfield  {author} {\bibinfo {author} {\bibfnamefont {V.}~\bibnamefont
  {Khachatryan}} \emph {et~al.} (\bibinfo {collaboration} {CMS}),\ }\href
  {\doibase 10.1007/JHEP04(2016)169} {\bibfield  {journal} {\bibinfo  {journal}
  {JHEP}\ }\textbf {\bibinfo {volume} {04}},\ \bibinfo {pages} {169} (\bibinfo
  {year} {2016})},\ \Eprint {http://arxiv.org/abs/1603.02248} {arXiv:1603.02248
  [hep-ex]} \BibitemShut {NoStop}%
\bibitem [{\citenamefont {del Aguila}\ \emph {et~al.}(2008)\citenamefont {del
  Aguila}, \citenamefont {de~Blas},\ and\ \citenamefont
  {Perez-Victoria}}]{delAguila:2008pw}%
  \BibitemOpen
  \bibfield  {author} {\bibinfo {author} {\bibfnamefont {F.}~\bibnamefont {del
  Aguila}}, \bibinfo {author} {\bibfnamefont {J.}~\bibnamefont {de~Blas}}, \
  and\ \bibinfo {author} {\bibfnamefont {M.}~\bibnamefont {Perez-Victoria}},\
  }\href {\doibase 10.1103/PhysRevD.78.013010} {\bibfield  {journal} {\bibinfo
  {journal} {Phys. Rev.}\ }\textbf {\bibinfo {volume} {D78}},\ \bibinfo {pages}
  {013010} (\bibinfo {year} {2008})},\ \Eprint {http://arxiv.org/abs/0803.4008}
  {arXiv:0803.4008 [hep-ph]} \BibitemShut {NoStop}%
\bibitem [{\citenamefont {Akhmedov}\ \emph {et~al.}(2013)\citenamefont
  {Akhmedov}, \citenamefont {Kartavtsev}, \citenamefont {Lindner},
  \citenamefont {Michaels},\ and\ \citenamefont {Smirnov}}]{Akhmedov:2013hec}%
  \BibitemOpen
  \bibfield  {author} {\bibinfo {author} {\bibfnamefont {E.}~\bibnamefont
  {Akhmedov}}, \bibinfo {author} {\bibfnamefont {A.}~\bibnamefont
  {Kartavtsev}}, \bibinfo {author} {\bibfnamefont {M.}~\bibnamefont {Lindner}},
  \bibinfo {author} {\bibfnamefont {L.}~\bibnamefont {Michaels}}, \ and\
  \bibinfo {author} {\bibfnamefont {J.}~\bibnamefont {Smirnov}},\ }\href
  {\doibase 10.1007/JHEP05(2013)081} {\bibfield  {journal} {\bibinfo  {journal}
  {JHEP}\ }\textbf {\bibinfo {volume} {05}},\ \bibinfo {pages} {081} (\bibinfo
  {year} {2013})},\ \Eprint {http://arxiv.org/abs/1302.1872} {arXiv:1302.1872
  [hep-ph]} \BibitemShut {NoStop}%
\bibitem [{\citenamefont {Basso}\ \emph {et~al.}(2014)\citenamefont {Basso},
  \citenamefont {Fischer},\ and\ \citenamefont {van~der Bij}}]{Basso:2013jka}%
  \BibitemOpen
  \bibfield  {author} {\bibinfo {author} {\bibfnamefont {L.}~\bibnamefont
  {Basso}}, \bibinfo {author} {\bibfnamefont {O.}~\bibnamefont {Fischer}}, \
  and\ \bibinfo {author} {\bibfnamefont {J.~J.}\ \bibnamefont {van~der Bij}},\
  }\href {\doibase 10.1209/0295-5075/105/11001} {\bibfield  {journal} {\bibinfo
   {journal} {Europhys. Lett.}\ }\textbf {\bibinfo {volume} {105}},\ \bibinfo
  {pages} {11001} (\bibinfo {year} {2014})},\ \Eprint
  {http://arxiv.org/abs/1310.2057} {arXiv:1310.2057 [hep-ph]} \BibitemShut
  {NoStop}%
\bibitem [{\citenamefont {Antusch}\ and\ \citenamefont
  {Fischer}(2014)}]{Antusch:2014woa}%
  \BibitemOpen
  \bibfield  {author} {\bibinfo {author} {\bibfnamefont {S.}~\bibnamefont
  {Antusch}}\ and\ \bibinfo {author} {\bibfnamefont {O.}~\bibnamefont
  {Fischer}},\ }\href {\doibase 10.1007/JHEP10(2014)094} {\bibfield  {journal}
  {\bibinfo  {journal} {JHEP}\ }\textbf {\bibinfo {volume} {10}},\ \bibinfo
  {pages} {094} (\bibinfo {year} {2014})},\ \Eprint
  {http://arxiv.org/abs/1407.6607} {arXiv:1407.6607 [hep-ph]} \BibitemShut
  {NoStop}%
\bibitem [{\citenamefont {Aaboud}\ \emph {et~al.}(2016)\citenamefont {Aaboud}
  \emph {et~al.}}]{Aaboud:2016qeg}%
  \BibitemOpen
  \bibfield  {author} {\bibinfo {author} {\bibfnamefont {M.}~\bibnamefont
  {Aaboud}} \emph {et~al.} (\bibinfo {collaboration} {ATLAS}),\ }\href
  {\doibase 10.1088/1367-2630/18/9/093016} {\bibfield  {journal} {\bibinfo
  {journal} {New J. Phys.}\ }\textbf {\bibinfo {volume} {18}},\ \bibinfo
  {pages} {093016} (\bibinfo {year} {2016})},\ \Eprint
  {http://arxiv.org/abs/1605.06035} {arXiv:1605.06035 [hep-ex]} \BibitemShut
  {NoStop}%
\bibitem [{\citenamefont {Sirunyan}\ \emph {et~al.}(2017)\citenamefont
  {Sirunyan} \emph {et~al.}}]{Sirunyan:2017yrk}%
  \BibitemOpen
  \bibfield  {author} {\bibinfo {author} {\bibfnamefont {A.~M.}\ \bibnamefont
  {Sirunyan}} \emph {et~al.} (\bibinfo {collaboration} {CMS}),\ }\href
  {\doibase 10.1007/JHEP07(2017)121} {\bibfield  {journal} {\bibinfo  {journal}
  {JHEP}\ }\textbf {\bibinfo {volume} {07}},\ \bibinfo {pages} {121} (\bibinfo
  {year} {2017})},\ \Eprint {http://arxiv.org/abs/1703.03995} {arXiv:1703.03995
  [hep-ex]} \BibitemShut {NoStop}%
\bibitem [{\citenamefont {Aad}\ \emph {et~al.}(2014)\citenamefont {Aad} \emph
  {et~al.}}]{Aad:2014pda}%
  \BibitemOpen
  \bibfield  {author} {\bibinfo {author} {\bibfnamefont {G.}~\bibnamefont
  {Aad}} \emph {et~al.} (\bibinfo {collaboration} {ATLAS}),\ }\href {\doibase
  10.1007/JHEP06(2014)035} {\bibfield  {journal} {\bibinfo  {journal} {JHEP}\
  }\textbf {\bibinfo {volume} {06}},\ \bibinfo {pages} {035} (\bibinfo {year}
  {2014})},\ \Eprint {http://arxiv.org/abs/1404.2500} {arXiv:1404.2500
  [hep-ex]} \BibitemShut {NoStop}%
\bibitem [{\citenamefont {Aaboud}\ \emph {et~al.}(2017)\citenamefont {Aaboud}
  \emph {et~al.}}]{Aaboud:2017dmy}%
  \BibitemOpen
  \bibfield  {author} {\bibinfo {author} {\bibfnamefont {M.}~\bibnamefont
  {Aaboud}} \emph {et~al.} (\bibinfo {collaboration} {ATLAS}),\ }\href@noop {}
  {\  (\bibinfo {year} {2017})},\ \Eprint {http://arxiv.org/abs/1706.03731}
  {arXiv:1706.03731 [hep-ex]} \BibitemShut {NoStop}%
\bibitem [{\citenamefont {Aad}\ \emph {et~al.}(2016)\citenamefont {Aad} \emph
  {et~al.}}]{Aad:2016tuk}%
  \BibitemOpen
  \bibfield  {author} {\bibinfo {author} {\bibfnamefont {G.}~\bibnamefont
  {Aad}} \emph {et~al.} (\bibinfo {collaboration} {ATLAS}),\ }\href {\doibase
  10.1140/epjc/s10052-016-4095-8} {\bibfield  {journal} {\bibinfo  {journal}
  {Eur. Phys. J.}\ }\textbf {\bibinfo {volume} {C76}},\ \bibinfo {pages} {259}
  (\bibinfo {year} {2016})},\ \Eprint {http://arxiv.org/abs/1602.09058}
  {arXiv:1602.09058 [hep-ex]} \BibitemShut {NoStop}%
\bibitem [{\citenamefont {Sjostrand}\ \emph {et~al.}(2006)\citenamefont
  {Sjostrand}, \citenamefont {Mrenna},\ and\ \citenamefont
  {Skands}}]{Sjostrand:2006za}%
  \BibitemOpen
  \bibfield  {author} {\bibinfo {author} {\bibfnamefont {T.}~\bibnamefont
  {Sjostrand}}, \bibinfo {author} {\bibfnamefont {S.}~\bibnamefont {Mrenna}}, \
  and\ \bibinfo {author} {\bibfnamefont {P.~Z.}\ \bibnamefont {Skands}},\
  }\href {\doibase 10.1088/1126-6708/2006/05/026} {\bibfield  {journal}
  {\bibinfo  {journal} {JHEP}\ }\textbf {\bibinfo {volume} {05}},\ \bibinfo
  {pages} {026} (\bibinfo {year} {2006})},\ \Eprint
  {http://arxiv.org/abs/hep-ph/0603175} {arXiv:hep-ph/0603175 [hep-ph]}
  \BibitemShut {NoStop}%
\bibitem [{\citenamefont {de~Favereau}\ \emph {et~al.}(2014)\citenamefont
  {de~Favereau}, \citenamefont {Delaere}, \citenamefont {Demin}, \citenamefont
  {Giammanco}, \citenamefont {Lemaître}, \citenamefont {Mertens},\ and\
  \citenamefont {Selvaggi}}]{deFavereau:2013fsa}%
  \BibitemOpen
  \bibfield  {author} {\bibinfo {author} {\bibfnamefont {J.}~\bibnamefont
  {de~Favereau}}, \bibinfo {author} {\bibfnamefont {C.}~\bibnamefont
  {Delaere}}, \bibinfo {author} {\bibfnamefont {P.}~\bibnamefont {Demin}},
  \bibinfo {author} {\bibfnamefont {A.}~\bibnamefont {Giammanco}}, \bibinfo
  {author} {\bibfnamefont {V.}~\bibnamefont {Lemaître}}, \bibinfo {author}
  {\bibfnamefont {A.}~\bibnamefont {Mertens}}, \ and\ \bibinfo {author}
  {\bibfnamefont {M.}~\bibnamefont {Selvaggi}} (\bibinfo {collaboration}
  {DELPHES 3}),\ }\href {\doibase 10.1007/JHEP02(2014)057} {\bibfield
  {journal} {\bibinfo  {journal} {JHEP}\ }\textbf {\bibinfo {volume} {02}},\
  \bibinfo {pages} {057} (\bibinfo {year} {2014})},\ \Eprint
  {http://arxiv.org/abs/1307.6346} {arXiv:1307.6346 [hep-ex]} \BibitemShut
  {NoStop}%
\bibitem [{\citenamefont {Selvaggi}(2014)}]{Selvaggi:2014mya}%
  \BibitemOpen
  \bibfield  {author} {\bibinfo {author} {\bibfnamefont {M.}~\bibnamefont
  {Selvaggi}},\ }\bibfield  {booktitle} {\emph {\bibinfo {booktitle}
  {{Proceedings, 15th International Workshop on Advanced Computing and Analysis
  Techniques in Physics Research (ACAT 2013)}}},\ }\href {\doibase
  10.1088/1742-6596/523/1/012033} {\bibfield  {journal} {\bibinfo  {journal}
  {J. Phys. Conf. Ser.}\ }\textbf {\bibinfo {volume} {523}},\ \bibinfo {pages}
  {012033} (\bibinfo {year} {2014})}\BibitemShut {NoStop}%
\bibitem [{\citenamefont {Mertens}(2015)}]{Mertens:2015kba}%
  \BibitemOpen
  \bibfield  {author} {\bibinfo {author} {\bibfnamefont {A.}~\bibnamefont
  {Mertens}},\ }\bibfield  {booktitle} {\emph {\bibinfo {booktitle}
  {{Proceedings, 16th International workshop on Advanced Computing and Analysis
  Techniques in physics (ACAT 14)}}},\ }\href {\doibase
  10.1088/1742-6596/608/1/012045} {\bibfield  {journal} {\bibinfo  {journal}
  {J. Phys. Conf. Ser.}\ }\textbf {\bibinfo {volume} {608}},\ \bibinfo {pages}
  {012045} (\bibinfo {year} {2015})}\BibitemShut {NoStop}%
\bibitem [{\citenamefont {Cacciari}\ \emph {et~al.}(2008)\citenamefont
  {Cacciari}, \citenamefont {Salam},\ and\ \citenamefont
  {Soyez}}]{Cacciari:2008gp}%
  \BibitemOpen
  \bibfield  {author} {\bibinfo {author} {\bibfnamefont {M.}~\bibnamefont
  {Cacciari}}, \bibinfo {author} {\bibfnamefont {G.~P.}\ \bibnamefont {Salam}},
  \ and\ \bibinfo {author} {\bibfnamefont {G.}~\bibnamefont {Soyez}},\ }\href
  {\doibase 10.1088/1126-6708/2008/04/063} {\bibfield  {journal} {\bibinfo
  {journal} {JHEP}\ }\textbf {\bibinfo {volume} {04}},\ \bibinfo {pages} {063}
  (\bibinfo {year} {2008})},\ \Eprint {http://arxiv.org/abs/0802.1189}
  {arXiv:0802.1189 [hep-ph]} \BibitemShut {NoStop}%
\bibitem [{\citenamefont {Cacciari}\ and\ \citenamefont
  {Salam}(2006)}]{Cacciari:2005hq}%
  \BibitemOpen
  \bibfield  {author} {\bibinfo {author} {\bibfnamefont {M.}~\bibnamefont
  {Cacciari}}\ and\ \bibinfo {author} {\bibfnamefont {G.~P.}\ \bibnamefont
  {Salam}},\ }\href {\doibase 10.1016/j.physletb.2006.08.037} {\bibfield
  {journal} {\bibinfo  {journal} {Phys. Lett.}\ }\textbf {\bibinfo {volume}
  {B641}},\ \bibinfo {pages} {57} (\bibinfo {year} {2006})},\ \Eprint
  {http://arxiv.org/abs/hep-ph/0512210} {arXiv:hep-ph/0512210 [hep-ph]}
  \BibitemShut {NoStop}%
\bibitem [{\citenamefont {Cacciari}\ \emph {et~al.}(2012)\citenamefont
  {Cacciari}, \citenamefont {Salam},\ and\ \citenamefont
  {Soyez}}]{Cacciari:2011ma}%
  \BibitemOpen
  \bibfield  {author} {\bibinfo {author} {\bibfnamefont {M.}~\bibnamefont
  {Cacciari}}, \bibinfo {author} {\bibfnamefont {G.~P.}\ \bibnamefont {Salam}},
  \ and\ \bibinfo {author} {\bibfnamefont {G.}~\bibnamefont {Soyez}},\ }\href
  {\doibase 10.1140/epjc/s10052-012-1896-2} {\bibfield  {journal} {\bibinfo
  {journal} {Eur. Phys. J.}\ }\textbf {\bibinfo {volume} {C72}},\ \bibinfo
  {pages} {1896} (\bibinfo {year} {2012})},\ \Eprint
  {http://arxiv.org/abs/1111.6097} {arXiv:1111.6097 [hep-ph]} \BibitemShut
  {NoStop}%
\bibitem [{ATL(2015)}]{ATLAS:2015-022}%
  \BibitemOpen
  \href
  {https://atlas.web.cern.ch/Atlas/GROUPS/PHYSICS/PUBNOTES/ATL-PHYS-PUB-2015-022/}
  {\emph {\bibinfo {title} {{Expected performance of the ATLAS b-tagging
  algorithms in Run-2}}}},\ \bibinfo {type} {Tech. Rep.}\ \bibinfo {number}
  {ATL-PHYS-PUB-2015-022}\ (\bibinfo  {institution} {CERN},\ \bibinfo {address}
  {Geneva},\ \bibinfo {year} {2015})\BibitemShut {NoStop}%
\bibitem [{\citenamefont {Drees}\ \emph {et~al.}(2015)\citenamefont {Drees},
  \citenamefont {Dreiner}, \citenamefont {Schmeier}, \citenamefont
  {Tattersall},\ and\ \citenamefont {Kim}}]{Drees:2013wra}%
  \BibitemOpen
  \bibfield  {author} {\bibinfo {author} {\bibfnamefont {M.}~\bibnamefont
  {Drees}}, \bibinfo {author} {\bibfnamefont {H.}~\bibnamefont {Dreiner}},
  \bibinfo {author} {\bibfnamefont {D.}~\bibnamefont {Schmeier}}, \bibinfo
  {author} {\bibfnamefont {J.}~\bibnamefont {Tattersall}}, \ and\ \bibinfo
  {author} {\bibfnamefont {J.~S.}\ \bibnamefont {Kim}},\ }\href {\doibase
  10.1016/j.cpc.2014.10.018} {\bibfield  {journal} {\bibinfo  {journal}
  {Comput. Phys. Commun.}\ }\textbf {\bibinfo {volume} {187}},\ \bibinfo
  {pages} {227} (\bibinfo {year} {2015})},\ \Eprint
  {http://arxiv.org/abs/1312.2591} {arXiv:1312.2591 [hep-ph]} \BibitemShut
  {NoStop}%
\bibitem [{\citenamefont {Dercks}\ \emph {et~al.}(2017)\citenamefont {Dercks},
  \citenamefont {Desai}, \citenamefont {Kim}, \citenamefont {Rolbiecki},
  \citenamefont {Tattersall},\ and\ \citenamefont {Weber}}]{Dercks:2016npn}%
  \BibitemOpen
  \bibfield  {author} {\bibinfo {author} {\bibfnamefont {D.}~\bibnamefont
  {Dercks}}, \bibinfo {author} {\bibfnamefont {N.}~\bibnamefont {Desai}},
  \bibinfo {author} {\bibfnamefont {J.~S.}\ \bibnamefont {Kim}}, \bibinfo
  {author} {\bibfnamefont {K.}~\bibnamefont {Rolbiecki}}, \bibinfo {author}
  {\bibfnamefont {J.}~\bibnamefont {Tattersall}}, \ and\ \bibinfo {author}
  {\bibfnamefont {T.}~\bibnamefont {Weber}},\ }\href {\doibase
  10.1016/j.cpc.2017.08.021} {\bibfield  {journal} {\bibinfo  {journal}
  {Comput. Phys. Commun.}\ }\textbf {\bibinfo {volume} {221}},\ \bibinfo
  {pages} {383} (\bibinfo {year} {2017})},\ \Eprint
  {http://arxiv.org/abs/1611.09856} {arXiv:1611.09856 [hep-ph]} \BibitemShut
  {NoStop}%
\bibitem [{\citenamefont {Read}(2002)}]{Read:2002hq}%
  \BibitemOpen
  \bibfield  {author} {\bibinfo {author} {\bibfnamefont {A.~L.}\ \bibnamefont
  {Read}},\ }\bibfield  {booktitle} {\emph {\bibinfo {booktitle} {{Advanced
  Statistical Techniques in Particle Physics. Proceedings, Conference, Durham,
  UK, March 18-22, 2002}}},\ }\href {\doibase 10.1088/0954-3899/28/10/313}
  {\bibfield  {journal} {\bibinfo  {journal} {J. Phys.}\ }\textbf {\bibinfo
  {volume} {G28}},\ \bibinfo {pages} {2693} (\bibinfo {year} {2002})},\
  \bibinfo {note} {[,11(2002)]}\BibitemShut {NoStop}%
\end{thebibliography}%
\end{document}